\documentclass[showpacs,amsmath,amssymb,amsfonts,a4paper,aps,prd]{revtex4}
\usepackage[colorlinks, linkcolor={RoyalBlue},citecolor={Red}]{hyperref}
\usepackage{graphicx,color}
\usepackage{float}
\usepackage[utf8]{inputenc}
\usepackage{multirow}

\usepackage[dvipsnames]{xcolor}

\begin{document}

\title{ Warm Higgs G-inflation: predictions and constraints from Planck 2015 likelihood }

\author{Meysam Motaharfar}
\email{mmotaharfar2000@gmail.com}
\author{Erfan Massaeli}
\email{erfan.massaeli@gmail.com}
\author{Hamid Reza Sepangi}
\email{hr-sepangi@sbu.ac.ir}
\affiliation{$^1$Department of Physics, Shahid Beheshti University, G. C., Evin,Tehran 19839, Iran}

\begin{abstract}
We reconsider a recently proposed warm G-inflation scenario in which the Galileon scalar field concurrently dissipates its kinetic energy as the radiation fluid throughout inflation and the universe smoothly enters into a radiation dominated era without going through the reheating phase. It is shown that the perturbed second-order Langevin equation can be nicely simplified and solved by defining the Galileon dissipation factor, $Q_{G} = \frac{Q}{\mathcal{B}}$, resulting in a power spectrum utilizing a Green function approach for the dissipative coefficient independent of temperature. However, for a dissipation coefficient depending on temperature, the perturbed inflaton and radiation field equations will be coupled in the high temperature regime. Therefore, the produced radiation backreacts on the power spectrum, modifying it with a growing mode function in the high dissipation regime. Finally, a model is proposed in which the standard  Higgs boson dissipates into light mediator fields, for instance, fermions with a linear temperature dependent dissipative coefficient which can act as inflaton, thanks to the Galileon-like non-linear derivative interaction. The generated primordial perturbations in the G-dominant regime is in excellent agreement with Planck 2015 likelihood+ TTTEEE+BAO at large scales despite its large self-coupling $\lambda \sim 0.13$ through accommodating many light mediator fields. However, although such primordial perturbations may also get amplified by several orders of magnitude at small scales due to the presence of the growing mode function,  warm G-inflation shows a striking feature in that the growing mode can be controlled or completely disappeared by decreasing the value of the propagating sound speed $c_{s}$.
\end{abstract}

\date{\today}

\maketitle

\section{Introduction}

The most recent cosmological observational data unanimously confirms that the universe is expanding, spatially flat, homogeneous and isotropic on large scales and the Large Scale Structure (LSS) is originated from the inhomogeneity in primordial scalar perturbations which are adiabatic Gaussian with quasi-invariant power spectrum \cite{\iffalse Observational data\fi Riess:1998cb, Komatsu:2010fb}. While from a theoretical point of view, \textit{inflation} \cite{\iffalse history of inflation\fi Starobinsky:1979ty, Linde:1984ir}, a finite period of quasi-de Sitter accelerated expansion phase where the energy density of universe is dominated by a slowly-evolving scalar field called ``\textit{inflaton}'', can set the initial conditions giving rise to a high degree of flatness and homogeneity, it can also serve as a casual mechanism to seed the acoustic peaks in Cosmic Microwave Background (CMB) Radiation as well as account for distribution of the LSS from evolution of primordial quantum vacuum fluctuations during inflation \cite{\iffalse inflation predictions\fi Linde:1981mu, Guth:1980zm, Guth:1982ec, Starobinsky:1982ee, Mukhanov:1981xt, Bardeen:1983qw}. Such a rapid expansion should typically be followed by a radiation dominated era allowing for synthesis of primordial nuclei, necessitating energy exchange between the inflaton and radiation fields by taking into account dissipation processes.

Dissipation processes determine how ultimately vacuum energy density, stored in the inflaton field, ends up converting to radiation, thus allowing the universe to  make a transition from inflationary phase to the radiation dominated era. Therefore, there are two pictures for dynamics of the inflation depending on how and when dissipation processes occur. The first and conventional one is the isentropic cold inflation (CI) where the inflaton field is isolated from interacting with other subdominant quantum fields, except gravitation, whereby the universe goes through a first order phase transition and its temperature drastically decreases. After such thermodynamically supercooling phase, the inflaton starts oscillating around the minimum of its potential and progressively dissipates its kinetic energy into other relativistic light degrees of freedom that thermalizes and provides the radiation bath required by the Standard Big Bang Cosmology (SBBC). This stage in which the universe heats up again is often called (p)reheating \cite{\iffalse reheating phase\fi Kofman:1994rk, Kofman:1997yn, Albrecht:1982mp}. In contrast to this picture, there is a mechanism dubbed non-isentropic warm inflation (WI) \cite{\iffalse warm inflation\fi Berera:1995wh, Berera:1995ie, Berera:1999} in which inflaton coexists with other sub-leading quantum fields where their coupling is strong enough so that their effects may not be neglected. Therefore, the state of inflationary universe is not a perfect vacuum state but rather an excited statistical state, with thermal state being the most examined, although vacuum energy is the dominant component for accelerated expansion to take place. Consequently, the dissipation process occurs not only after but also during the slow-roll phase of inflation whereby a quasi-equilibrium thermal radiation bath is concurrently generated throughout inflation, compensating the supercooling phase observed in CI where radiation smoothly becomes the dominant ingredient of the universe after the inflationary expansion.

The dissipation process during inflation modifies both the homogeneous evolution of inflaton and the inhomogeneous fluctuations which bring about interesting predictions in comparison to the CI picture. The energy exchange between the inflaton field and radiation field is supplemented by dissipative coefficient $\Gamma$ which is translated into a supplementary damping viscous term $\Gamma \dot\phi$ in Klein-Gordon equation \cite{\iffalse Dissipation as friction term \fi Berera:2009qc, Bartrum:2013}. Therefore, such a damping friction term not only keeps primeval radiation fluid from being diluted but slows down the evolution of the inflaton field whereby inflation with steeper potentials may last for  prolonged periods. Consequently, the tensor-to-scalar ratio may be suppressed due to smaller energy density or a higher height for potential at the Hubble crossing time. On the other hand, the source of primordial density fluctuations stems from thermal fluctuations \cite{\iffalse thermal fluctuations\fi Yokoyama:1998ju, Taylor:2000ze, DeOliveira:2001he, DeOliveira:2002wk, Hall:2003zp, Moss:2008yb, Bartrum:2013oka} in a radiation bath which is transferred to the inflaton field as adiabatic curvature perturbations while quantum fluctuations of the inflaton field are the source of density perturbations in CI scenario.

The backreaction of thermal fluctuations is modulated by including a stochastic noise term leading to Stochastic Langevin Equation (SLE) where the energy exchange between the inflaton and radiation fields is also included by the factor $\Gamma\dot\phi$. This SLE which governs the evolution of thermal fluctuations is derived from first principles in quantum field theory by applying the equivalence principle to non-expanding results \cite{\iffalse Langevin equation\fi Berera:1996nv, Berera:2005f, Ramos:2013, Bastero:2011s, Bastero:2014d, Visinelli}. However, when the dissipative coefficient depends on temperature with a positive power,  the thermal fluctuations in the high-temperature regime not only come from the stochastic noise term but also from the coupling between the perturbed inflaton and radiation field equations due to the presence of $\Gamma \dot\phi$. It was first shown in \cite{\iffalse Coupling equation in WI\fi Graham:2009}  that the power spectrum is enhanced due to the backreaction of produced radiation fluid by a growing mode function in contrast to previous works.

The specific functional form of the dissipative coefficient when considering both the inflaton field and temperature of the radiation bath, which is derived from first principles in quantum field theory in adiabatic approximation, depends on how the inflaton field decays into light degrees of freedom during inflation. For instance, inflaton may decay into light degrees of freedom through a two-stage mechanism where dissipation is originated from the coupling of the inflaton and a massive mediator, subsequently decaying into light degrees of freedom. As it has been shown in \cite{\iffalse two-stage mechanism\fi Berera:2008ar}, such a two-stage interaction configuration may result in a large enough dissipation while thermal corrections to the inflaton potential arising from this mechanism do not spoil the flatness of the potential. Therefore, the difficulty in producing viable warm inflationary models does not exist. Utilizing this mechanism, the specific form of the dissipative coefficient in the context of close-to-equilibrium approximation has a power-law temperature dependence which depending on the mass of the mediator,  can be cubic, linear or even inverse in temperature, depending on being in a low or high temperature regime \cite{\iffalse the form of dissipation\fi Bastero:2011, Bastero:2013, Moss:2006, Berera:2003}. It has however been demonstrated that the inverse temperature model produces a large thermal correction to the inflaton potential, precluding the WI scenario from taking place.

Summing up, the WI picture not only inherits the features of CI picture but also removes or alleviates some difficulties which exist in conventional CI scenario. First, it does not suffer from discrepancy coming  from matching two isolated stages; the inflationary period and subsequent reheating phase by adopting a smooth transition to the radiation dominated era at the end of inflation. While single field models of inflation are not embedded within UV-completion of Standard Model (SM) such as supergravity or String theory due to the so-called ``$\eta$-problem'', the WI scenario circumvents this problem by introducing the dissipative coefficient whereby $\eta$ may acquire a large value \cite{\iffalse eta problem\fi Berera:2004vm}. Furthermore, it can alleviate the initial condition \cite{\iffalse initial condition\fi Berera:2000xz}, cure overlarge amplitude of the inflaton field and contributes a very appealing mechanism for baryogenesis where spontaneous lepto/baryogenesis may easily be realized \cite{\iffalse baryogenesis in WI\fi BasteroGil:2011cx, Brandenberger:2003kc}.

Although there are many indications of physics beyond the SM, so far no direct evidence has been reported for neutrino oscillations \cite{\iffalse neutrino oscillation\fi Gonzalez:2008}. Hence, the only known scalar field responsible for inflation is the SM Higgs boson. However, Higgs-driven inflationary models with a renormalizeable self-interaction potential \cite{\iffalse chaotic\fi Linde:1983gd} cannot be responsible for inflation as long as its kinetic term is canonical and minimally coupled to gravity, since it produces too large curvature and tensor perturbations which are not consistent with current observed universe \cite{Ade:2015xua} due to strong self-interaction of SM Higgs boson. Therefore, to survive Higgs-driven inflation confronting with observations, several variant Higgs-driven inflationary models have been proposed by imposing distinct modifications to the effective Lagrangian in order to suppress the energy scale of inflation including a non-minimally coupled term to gravity with a large coupling \cite{Barvinsky:2008ia}, non-minimal field derivative coupling with Einstein tensor (new Higgs inflation) \cite{Germani:2010gm, Germani:2010ux}, non-standard higher order kinetic term, dubbed k-inflation (running kinetic inflation) \cite{ArmendarizPicon:1999rj, Garriga:1999vw} such as ghost condensate \cite{ArkaniHamed:2003uz} and Dirac-Born-Infeld inflationary models \cite{Alishahiha:2004eh}. Except for the first where the amplitude of curvature perturbations is suppressed due to the large effective Planck scale, others are kinetically modified, resulting in extra viscous terms in the inflaton dynamical equations whereby the evolution of the inflaton field may be slowed down (slotheon mechanism \cite{Germani}); therefore, inflationary phase lasts for a longer period and becomes consistent with observations even for strong self-coupling and steeper potentials.

One way to kinetically modify the effective Lagrangian is by incorporating higher order non-standard kinetic terms which usually result in a new degree of freedom followed by unwanted ghost instabilities; therefore, having a Lagrangian with a higher derivative term of the scalar field which does not lead to a new degree of freedom is desirable. Currently, it has been demonstrated that a particular combination of higher derivative kinetic terms not only maintains both the scalar and gravitational field equations to second order but also does not lead to a new degree of freedom \cite{Deffayet:2009wt, Deffayet:2009mn}. The scalar field having such properties is known as the Galileon since it possesses a Galileon shift symmetry in the limit of  Minkowski space-time. Such a scalar field has initially been investigated in the context of modified gravity and dark energy \cite{Chow:2009fm, Silva:2009km}. Recently, it has been shown that the scalar field with a Galileon interaction term can violate the null energy condition stably, motivating authors in \cite{\iffalse G-inflation\fi Kobayashi:2010cm} to propose a Galileon driven inflationary model dubbed G-inflation including canonical and non-canonical scalar field models of inflation which is literally known as kinetic gravity braiding models. The striking characteristics of such inflationary models are that they can produce the scale-invariant spectral index even in an exactly de-Sitter background and  tensor-to-scalar ratio can take larger values than that in conventional inflation due to violation of the consistency relation, that is $r = - 8.7 n_{t}$. Other predictions of G-inflation have been explored in \cite{\iffalse G-inflation predictions\fi Kamada:2010qe, Ohashi:2012wf, Kamada:2013bia}

Although Higgs G-inflation  \cite{Kamada:2010qe} is consistent with observations even for large self-couplings,  roughly around $0.13$ from quantum field theory point of view,  it has very recently been realized that it suffers from the absence of an oscillatory phase typically accompanied by a negatively squared propagating sound speed leading to a Laplace equation for curvature perturbations instead of a wave equation, while producing unstable small-scale perturbations \cite{Ohashi:2012wf}. To resolve the problem, the authors in \cite{Kamada:2013bia} have added an extra quadratic non-standard kinetic term to the action in order to obtain positive sound speed resulting in the required reheating phase despite large self-coupling of the Higgs self-interaction potential. Yet, it has been shown that G-inflation in the WI picture not only survives but also simultaneously inherits the attractive features of WI picture presented in \cite{Motaharfar:2017dxh} (for a review of   warm inflationary models see \cite{\iffalse warm inflationary models \fi Bastero-Gil:2016qru, Richa:2018, Bastero-Gil:2018, Bastero:2015, Benetti:2017, Visinelli:2016, Cai:2011, Mishra:2012, Panotopoulos:2015, Zhang:2014, Herrera:2017}). However, it should be noted that the calculation of the power spectrum in \cite{Motaharfar:2017dxh} has been done by assuming that temperature fluctuations are very small or in a weak dissipation regime. In fact, we have not considered the coupling between the perturbed inflaton and radiation fields in a strong dissipation regime. Thus, the resultant power spectrum is just consistent in the weak dissipation regime or for dissipative coefficient which is only a function of the inflaton field.

Having the above points in mind, the goal of this paper is twofold. First, computing the enhanced power spectrum including the effect of coupling between the perturbed inflaton and radiation fields for temperature dependence of a dissipative coefficient in the strong regime and second, to investigate how such modification may impact the constraints on the parameters of the model when confronting with the Planck 2015 data. Therefore, the layout of the paper is the following. First, we review the background equations of Warm G-inflation (WGI) and summarize the conditions obtained to validate slow-roll regime in WGI  \cite{Motaharfar:2017dxh}. In section \ref{Primordial power spectrum in WGI}, we derive the power spectrum for WGI by utilizing the newly defined Galileon dissipation factor and Green function approach extended in \cite{Graham:2009} for both temperature independent and the dependent function of dissipative coefficient and illustrate how the previous results are improved. Next, we investigate the theoretical predictions obtained due to improved power spectrum in section \ref{Observable quantities and predictions}. We solve the model for the linear temperature dependence of dissipative coefficient, Higgs self-interaction potential and the general form of $G(\phi, X)$ and obtain all dynamical parameters and observable quantities based on Galileon dissipation factor $Q_{G}$ in section \ref{Warm Higgs G-inflation}. In section \ref{Observable quantities and constraints}, we explain how to constrain the model utilizing results derived in the previous section. Finally, conclusions are drawn and possible future projects discussed in the last two section. Throughout the paper, we adopt the metric signature $(- , +, +, +)$ and choose units so that $c= \hbar=1$.

\section{Warm G-inflation: dynamics}
In a WI scenario, the inflaton dissipates its vacuum energy to other quantum fields during inflation and if such dissipation produces light degrees of freedom which thermalize within a Hubble time, then the radiation fluid is concurrently produced and continually replenished by the effective decay of the inflaton field. Hence, the energy exchange between the inflaton and radiation fields in the leading adiabatic approximation in a spatially flat, homogeneous and isotropic universe with the expansion rate $H$ reading \cite{Benetti:2017}
\begin{align}
\dot\rho_{\phi}+ 3H\left(\rho_{\phi}+ P_{\phi}\right)& = - \Gamma \dot\phi^{2},\label{a3}\\
\dot \rho_{R} + 4H \rho_{R} &=  \Gamma \dot \phi^{2}\label{p1},
\end{align}
where $\rho_{R} = \frac{\pi^{2}}{30} g_{\star}(T)T^{4}$ with $g_{\star}$ being the relativistic degree of freedom and $T$ is the universal temperature. Furthermore, a dot denotes time derivative, $H = \frac{\dot a(t)}{a(t)}$ with $a(t)$ being the scale factor as a function of cosmic time $t$, $\phi$ is the homogeneous inflaton field as a function of cosmic time, $\Gamma(\phi, T)$ is the dissipative coefficient as a function of both $\phi$ and $T$, $P_{\phi}$ represents the inflaton pressure, and $\rho_{\phi}$ and $\rho_{R}$ are the energy density of the inflaton and radiation fields respectively. Moreover, the Hubble parameter is related to the total energy density and pressure through the following gravitational equations

\begin{align}
3M^{2}_{pl}H^{2}&= \rho, \label{p2}\\
-M^{2}_{pl}(3H^{2}+2\dot H) &= P,
\end{align}
where $\rho$ and $P$ are total energy density and pressure containing both the inflaton and radiation contributions respectively and $M_{pl} = (8\pi G_{N})^{-\frac{1}{2}} = 2.44\times 10^{18} Gev$ is the reduced Planck mass with $G_{N}$ being the gravitational constant.

To reconstruct G-inflation in the context of WI scenario, we proceed by considering the following multicomponent, kinetically modified, minimally coupled  Lagrangian as follows

\begin{align}\label{x1}
\mathcal{L} = \frac{M^{2}_{pl}}{2}R + X- V(\phi, T) - G(\phi, X) \box \phi + \mathcal{L}_{R} + \mathcal{L}_{int},
\end{align}
where $R$ is Ricci scalar, $X  = - \frac{1}{2} g^{\mu\nu} \partial_{\mu}\phi \partial_{\nu} \phi$ is the standard kinetic term with $g^{\mu\nu}$ being the four dimensional FLRW metric, $V(\phi, T)$ is the potential energy density as a function of $\phi$ and $T$, $G(\phi, X)$ represents an arbitrary function of $\phi$ with $X$, $\mathcal{L}_{R}$ and $\mathcal{L}_{int}$ denoting the Lagrangian for radiation field and interaction terms between inflaton and other subdominant fields, respectively. Therefore, $\rho_{\phi}$ and $P_{\phi}$ for the above Lagrangian are given by \cite{Kobayashi:2010cm}

\begin{align}
\rho_{\phi}& = X+V(\phi, T)+ 6HG_{,X}X\dot\phi-2G_{,\phi}X,\label{a1}\\
P_{\phi}&= X-V(\phi, T) - 2\left(G_{,\phi}+G_{,X}\ddot\phi\right)X\label{a2},
\end{align}
with $G_{,\phi} = \frac{\partial G}{\partial \phi}$. Inserting (\ref{a1}, \ref{a2}) into Eq. (\ref{a3}), one can find the equivalent modified Klein-Gordon equation as follows \cite{Motaharfar:2017dxh}

\begin{align}\label{d1}
\mathcal{B}\ddot\phi(t) + 3 H \mathcal{A}\dot \phi(t)+V_{,\phi}(\phi, T)=0,
\end{align}
with

\begin{align}
\mathcal{A}= &1+Q+ 3H\dot\phi G_{,X}+ \frac{\dot H \dot\phi G_{,X}}{H}- 2G_{,\phi}+2XG_{,\phi X}- \frac{G_{,\phi\phi}\dot\phi}{3H},\\
\mathcal{B}= &1+6H\dot\phi G_{,X}+ 6H \dot\phi XG_{,XX}-2G_{,\phi}-2XG_{,\phi X},
\end{align}
where $Q$ which measures the effectiveness of dissipation process is defined as

\begin{align}
Q = \frac{\Gamma}{3H}.
\end{align}
Looking at Eq. (\ref{d1}), one may easily find that the damping coefficient $\Gamma$ and Galileon interaction term have appeared as an additional viscous terms in the inflaton equation of motion whereby further novel inflationary models with steeper potentials may also be included in WGI due to the presence of these two effects simultaneously.

The duration of inflation can be measured by the number of e-folding which is defined as

\begin{align}\label{az}
N = \int^{t_{end}}_{t_{hc}} H d t = \int^{\phi_{end}}_{\phi_{hc}} \frac{H}{\dot\phi} d \phi,
\end{align}
where subscripts "$end$" and "$hc$" represent the end of inflation and Hubble crossing time, respectively. To resolve the horizon problem, inflation should be proceeded by a sufficient number of e-folding; therefore, the slow-roll condition $|\epsilon_{Y}|\ll 1$ where $\epsilon_{Y} = \frac{d\ln Y}{d N}$ with $Y$ being any of background quantities, should be satisfied.  In fact, slow-roll regime implies that energy of the universe is dominated by the potential ($\dot H\ll H^{2}\sim V$), the inflaton is slowly evolving ($\ddot \phi \ll H\dot \phi$) and radiation is quasi-statically produced ($\dot \rho_{R}\ll 4 H \rho_{R}$). Thus, inflation takes place when the condition $|\epsilon_{H}|<1$ is satisfied, implying $\ddot a>0$ and will terminate where slow-roll condition is violated first ($\epsilon_{H} = 1$).  Consequently, Eqs. (\ref{p1}, \ref{p2}, \ref{d1}) reduce to \cite{Motaharfar:2017dxh}
\begin{align}
3 M^{2}_{pl} H^{2} & \simeq V\label{r},\\
3 H \mathcal{A} \dot \phi+ V_{,\phi} & \simeq 0 \label{e1},\\
\rho_{R} &\simeq \frac{3}{4}Q \dot \phi^{2} \label{e2},
\end{align}
where "$\simeq$"  represents an equality that holds in the slow-roll regime, and
\begin{align}
\mathcal{A}& \simeq 1+Q+ 3H\dot\phi G_{,X},\\
\mathcal{B}&\simeq 1+6H\dot\phi G_{,X}+ 6H \dot\phi XG_{,XX}.
\end{align}
For simplicity we define the effective Galileon dissipation factor as
\begin{align}
Q_{G} \equiv \frac{Q}{\mathcal{B}}.
\end{align}
Therefore, we can rewrite Eqs. (\ref{e1}, \ref{e2}) in terms of $Q_{G}$
\begin{align}
3 H \mathcal{B} \dot \phi \left(Q_{G} + \frac{\delta_{X}+ 3\delta_{GX}}{\delta_{X}+ 6(\kappa_{X}+1)\delta_{GX}}\right)+ V_{,\phi} & \simeq 0,\\
\rho_{R} &\simeq \frac{3}{4}\mathcal{B}Q_{G} \dot \phi^{2}.\label{t1}
\end{align}
In the $G$-dominant regime where $|\delta_{X}| \ll |\delta_{GX}|$ we have
\begin{align}\label{t2}
3 H \mathcal{B} \dot \phi \left(Q_{G} + \frac{1}{ 2(\kappa_{X}+1)}\right)+ V_{,\phi} & \simeq 0,
\end{align}
where $\kappa_{X} = \frac{XG_{,XX}}{G,X}$. Consequently, utilizing Eqs. (\ref{t1} and \ref{t2}), we may obtain parameters of the models such as $\phi, \dot\phi, T$ and the number of e-folding as a function of $Q_{G}$.

The validity of slow-roll approximations used to derive Eqs. (\ref{r}, \ref{e1} and \ref{e2}) depends on the Potential Slow-Roll (PSR) parameters
\begin{align*}
\epsilon \equiv \frac{M^{2}_{pl}}{2} \left(\frac{V_{,\phi}}{V}\right)^{2}, \ \ \ \ \  \eta \equiv M^{2}_{pl}\frac{V_{,\phi\phi}}{V},  \ \ \ \ \ \beta \equiv M^{2}_{pl}\frac{V_{,\phi}\Gamma_{,\phi}}{V\Gamma},
\end{align*}
which are supplemented by two more parameters, namely
\begin{align}
b \equiv \frac{T V_{,\phi T}}{V_{,\phi}}, \ \ \ \ \  c \equiv \frac{T \Gamma_{,T}}{\Gamma},
\end{align}
gauging the temperature dependence of the potential and damping coefficient, respectively. There are also three dimensionless parameters as follows
\begin{align}
\delta_{X} =  \frac{X}{M^{2}_{pl}H^{2}}, \ \ \ \ \ \delta_{GX} = \frac{\dot\phi X G_{,X}}{M^{2}_{pl} H},  \ \ \ \ \ \ \delta_{G\phi} = \frac{XG_{,\phi}}{M^{2}_{pl}H^{2}}.
\end{align}
Slow-roll approximations impose the following conditions on slow-roll parameters, for details see \cite{Motaharfar:2017dxh}
\begin{align}
\left\{|\epsilon|, |\eta|, |\beta| \right\}\ll \mathcal{A}, \ \ \ \left\{|\delta_{X}|, |\delta_{GX}|, |\delta_{G\phi}|\right\}\ll1, \ \ \  0< b\ll \frac{Q}{A}, \ \ \  |c|\le 4, \ \ \ \left|G_{,\phi}\right|=\left|\frac{\delta_{G\phi}}{\delta_{X}}\right|\ll 1.
\end{align}
Thus, $|\epsilon|, |\eta|$ may acquire large values in the strong dissipation regime alleviating the need for very flat potentials. Also, the condition on $b$ implies that thermal corrections to the potential are very small and that the Galileon interaction term is kinetically dominant. Furthermore, it deserves noting that the last condition in $G$-dominant regime reduces to $|\frac{\delta_{GX}}{\delta_{G\phi}}|\ll1$.

In the slow-roll regime, the relation between radiation and inflaton energy density utilizing Eqs. (\ref{e1}, \ref{e2}) is given by
\begin{align}\label{rr1}
\frac{\rho_{R}}{\rho_{\phi}} \simeq \frac{1}{2} \frac{Q\epsilon}{\mathcal{A}^{2}}.
\end{align}
Therefore, during inflation the energy density associated with the inflaton field predominates the radiation field ($\rho_{\phi}\gg \rho_{R}$) or in other words, radiation is somewhat suppressed as we expected. Although, radiation energy density in comparison to inflaton energy density is so small, it can be larger than the Hubble rate with $\rho_{R}^{\frac{1}{4}}> H$. Assuming thermalization, it can be roughly translated to the condition $T>H$ for which warm inflation occurs. Although, this condition may be satisfied for weak dissipation, at the end of inflation for $\epsilon \sim \mathcal{A}$ and $\frac{\rho_{R}}{\rho_{\phi}} \simeq \frac{1}{2} \frac{Q}{\mathcal{A}}$, radiation may be the dominant component for strong dissipation; consequently, the universe smoothly enter into a radiation dominated era without the reheating phase.

\section{Warm G-inflation: primordial power spectrum} \label{Primordial power spectrum in WGI}

As is well known, the prime characteristic of dissipating inflationary models which distinguishes them from the so-called cold inflation is that the nature of density fluctuations is due to thermal fluctuation in radiation field rather than quantum fluctuations. Thermal noise is transferred to the inflaton field mostly on small scales. As the comoving wavelength of perturbation expands, the thermal effects decrease until the fluctuations amplitude freezes out. This may occur when the wavelength of the fluctuation is still small in comparison to cosmological scales. These thermal fluctuations in radiation field are coupled to the inflaton field through the presence of damping terms in dynamical equations of inflation and their amplitude is fixed by the fluctuation-dissipation theorem \cite{Moss:2008yb,Hall:2003zp}. Hence, the interaction between the inflaton field and radiation field can be analyzed within the Schwinger-Keldysh approach in non-equilibrium field theory. In a flat spacetime, the field can be described by a stochastic system whose evolution is determined by Langevin equation. Consequently, evolution of the inflaton field in expanding universe is obtained by applying equivalence principle to non-expanding universe, replacing ordinary derivatives with covariant derivatives, leading to modified SLE after introducing a thermal  stochastic noise term
\begin{align}\label{SLE}
\bold{B}\ddot\Phi(x,t) + 3 H \bold{A} \dot\Phi(x,t)  - \bold{F} \frac{\nabla^{2}}{a^{2}}\Phi(x,t) + V_{,\Phi}=  \xi(x,t),
\end{align}
where
\begin{align}
\bold{A}= &1+Q+ 3H\dot\Phi G_{,X}+ \frac{\dot H \dot\Phi G_{,X}}{H}- 2G_{,\Phi}+2XG_{,\Phi X}- \frac{G_{,\Phi\Phi}\dot\Phi}{3H},\\
\bold{B}= &1+6H\dot\Phi G_{,X}+ 6H \dot\Phi XG_{,XX}-2G_{,\Phi}-2XG_{,\Phi X}-2\left(G_{,X}+XG_{,XX}\right)\frac{\nabla^{2}}{a^{2}}\Phi, \\
\bold{F}= & 1- 2 G_{,\Phi} + 2XG_{,\Phi X} + 4 H \dot\Phi G_{,X},
\end{align}
with $\xi$ being stochastic noise term. To obtain the evolution of thermal fluctuations, we expand $\Phi(x, t)$ around its background as $\Phi(x,t) = \phi(t) + \delta\phi(x,t)$ with $\phi(t)$ being the homogeneous field background and $\delta\phi$ a small field perturbation, that is $\delta\phi
\ll \phi(t)$. Therefore, we have dropped some terms in Eq. (\ref{SLE}) which results in second or higher order perturbations, see appendix \ref{we} for the complete form of SLE. Thus, the evolution of thermal density fluctuations in WGI, utilizing slow-roll regime and in Fourier space can be obtained from the following second-order Langevin equation, see appendix \ref{we} for details
\begin{align} \label{m1}
\mathcal{B}\delta\ddot\phi(\bold{k}, t)+3H\left(Q+\mathcal{B}\right)\delta\dot\phi(\bold{k},t)+\left(
c_{s}^{2}k^2a^{-2}\mathcal{B}-3H^{2}Q \frac{\beta}{\mathcal{A}} +V_{,\phi\phi}\right)\delta\phi(\bold{k},t)+3cH^2Q(\Gamma\dot\phi)^{-1}\delta\rho_R=\xi(\bold{k},t),
\end{align}
where $\bold{k}$ is the comoving coordinate wave-vector with $k = |\bold{k}|$ and we have used the following expression
\begin{align}
3H\delta Q \dot\phi=\delta \Gamma \dot\phi=3cH^2Q(\Gamma\dot\phi)^{-1}\delta\rho_R- 3 H^{2} Q \frac{\beta}{\mathcal{A}} \delta\phi,
\end{align}
with $c_{s}$ being the effective propagating speed of sound given by
\begin{align}
c^{2}_{s}=& \frac{\delta_{X}+4\delta_{GX}}{\delta_{X}+6 (\kappa_{X}+1) \delta_{GX}},
\end{align}
which a $G$-dominant regime ($|\delta_{GX}| \gg |\delta_{X}|$) becomes
\begin{align}
c^{2}_{s} = \frac{2}{3}\frac{G_{,X}}{G_{,X}+ XG_{,XX}}.
\end{align}
A look at Eq. ($\ref{m1}$) shows that the evolution of inflaton fluctuations depends on radiation fluctuations and  one should take it into account during calculating the inflaton fluctuations.

There are two sources of radiation fluctuations: first, it may come from purely statistical and caused by microphysics particle physics, as shown in reference \cite{Graham:2009}. Such fluctuations are subdominant in comparison to inflaton fluctuations and can be neglected during calculations. Second, they may arise from a momentum flux and energy flux transfer to radiation which plays the dominant role and is given by a second-order equation for radiation fluctuations in Fourier space as follows, see \cite{Graham:2009}
\begin{align}\label{h2}
\delta\ddot\rho_{r}+(9-c)H\delta\dot\rho_{r}+
\left((20-5c)H^2+{1\over 3}k^2a^{-2}\right)\delta\rho_{r}=- k^2a^{-2}(\Gamma\dot\phi)\,\delta\phi.
\end{align}
Consequently, the inflaton and radiation fluctuations are coupled together as appear in Eqs. ($\ref{m1}, \ref{h2}$) for $c\neq 0$ and $Q_{G}\gg 1$ and become decoupled for $c=0$  or for $Q_{G}\ll 1$. Therefore, we compute the inflaton power spectrum using Green functions first introduced in \cite{Graham:2009} for these cases in the next two sections.

\subsection{Case $c=0$ or $Q_{G}\ll 1$}
A glance at the Langevin equation $(\ref{m1})$ reveals its complicated structure, making it hopelessly difficult to achieve a solution. However, by introducing a new time coordinate $z= \frac{c_{s}k}{aH}$ and using the newly defined Galileon dissipation factor $Q_{G}$, we may rewrite Eq. ($\ref{m1}$) as follows
\begin{align}\label{m2}
 (1-\epsilon_{z})^{2} \delta\phi^{\prime\prime} - ((1-\epsilon_{z})\left(3Q_G+2 + \delta_{cs}) - \epsilon_{z} \eta_{z}\right)z^{-1}\delta\phi^{\prime} + \left(1+ {\frac{3\eta}{{z^{2}\mathcal{B}}} - \frac{3 Q_{G}\beta}{{z^{2}\mathcal{A}}}} \right) \delta\phi +3c z^{-2}Q_{G}(\Gamma\dot\phi)^{-1}\delta\rho_R=\frac{\xi}{ (Hz)^{2} \mathcal{B}},
\end{align}
where a prime denotes derivative with respect to $z$ and $\eta_{Y} = \frac{\dot \epsilon_{Y}}{H \epsilon_{Y}}$ is a second order slow-roll parameter with $\epsilon_{Y}$ being any first order slow-roll parameter. To obtain $(\ref{m2})$, we have used slow-roll parameters and the following relations
\begin{align}
\dot{z}&= -zH \left(1- \epsilon_{z}\right),~~\epsilon_{z} = \epsilon_{H}+ \delta_{c_{s}},\\
\ddot z&=zH^2\left((1-\epsilon_z)(1- \delta_{cs}) + \epsilon_{z} \eta_{z}\right),\\
\delta_{c_s}&\equiv \frac{\dot c_{s}}{Hc_{s}}=\frac{(3\kappa_X+1)\delta_{GX}}{c_s^2 \mathcal{B}^2 \delta_X }(\eta_X+\eta_{GX}).
\end{align}
In the slow-roll regime, we can drop the first order slow-roll parameters of coefficients in Eq. (\ref{m2}) in order to keep the Langevin equation first order in perturbations, which reduces to
\begin{align}\label{eq14}
\delta\phi^{\prime\prime}-(3Q_{G}+2)z^{-1}\delta\phi'
+\delta\phi
=(Hz)^{-2} \mathcal{B}^{-1}\xi,
\end{align}
where we have also dropped the term containing radiation field perturbations either by assuming that the dissipative coefficient is independent of temperature $(c = 0)$ or the dissipation regime is weak ($Q_{G}\ll 1$), implying that perturbations are assumed to be small and may be ignored. The above equation can be solved using Green function techniques, resulting in the following solution
\begin{align}\label{k1}
\delta{\phi}(\bold{k}, z) = H^{-2}\mathcal{B}^{-1 }\int_{z}^{\infty} d z^{\prime} G(z, z^{\prime}){(z^{\prime})^{-1-2\nu}} \xi,
\end{align}
where $G(z, z^{\prime})$ is the retarded Green function given by
\begin{align}
G(z, z^{\prime}) = \frac{\pi}{2} z^{\nu} {z^{\prime}}^{\nu} \left[J_{\nu}(z)Y_{\nu}(z^{\prime})- J_{\nu}(z^{\prime})Y_{\nu}(z)\right],
\end{align}
with $z<z^{\prime}$ and
\begin{align}
\nu \simeq \frac{3\left(1+Q_{G}\right)}{2} = \frac{\Gamma + 3 H \mathcal{B}}{2 H \mathcal{B}}.
\end{align}
The inflaton power spectrum $\mathcal{P}_{\phi}$ is defined as
\begin{align}
\mathcal{P}_{\phi}(k, z)  = \frac{k^{3}}{2\pi^{2}} \int \frac{d^{3} k^{\prime}}{(2\pi)^{3}} \langle \delta\phi(\bold{k}, z),\delta\phi(\bold{k}^{\prime}, z)\rangle.
\end{align}
Utilizing Eq.($\ref{k1}$), this can be written as

\begin{align}
\mathcal{P}_{\phi}(\bold{k}, z) &=  \frac{k^{3}}{2\pi^{2}H^{4}\mathcal{B}^{2}} \int \frac{d^{3} k^{\prime}}{(2\pi)^{3}}\int_{z}^{\infty} d z^{\prime}\int_{z}^{\infty} d z^{\prime\prime} G(z, z^{\prime}) G(z, z^{\prime\prime}) (z^{\prime}z^{\prime\prime})^{-1-2\nu}  \langle{\xi}(\bold{k}, z^{\prime}), {\xi}(\bold{k}^{\prime}, z^{\prime\prime}) \rangle.
\end{align}
The correlation function for the stochastic term may be re-scaled \cite{Ramos:2013} and written as
\begin{align}
\langle{\xi}(\bold{k}, z), {\xi}(\bold{k}^{\prime}, z^{\prime}) \rangle =  \frac{ (2\pi)^{3}H^{4} z^{3} z^{\prime}}{k^{2} {k^{\prime}}} \left(2 \Gamma_{eff} T\right) \delta(z-z^{\prime}) \delta^{3}(\bold{k}+\bold{k}^{\prime}).
\end{align}
Consequently, the power spectrum can be obtained from the following integration, see appendix \cite{Graham:2009} for this type of integrals
\begin{align}\label{p12}
\mathcal{P}_{\phi} = \frac{\Gamma_{eff}T}{\pi^{2}\mathcal{B}^{2}}\int_{z}^{\infty}dz^{\prime} G(z, z^{\prime})^{2} (z^\prime)^{2-4\nu},
\end{align}
where $\Gamma_{eff}$ is obtained by matching the power spectrum in flat space thermal field theory and that coming from Eq. ($\ref{p12}$) for large $z$ approximation \cite{Graham:2009}
\begin{align}
\Gamma_{eff} = \mathcal{B}^{2} c_{s}^{-1}\left(\frac{\Gamma}{\mathcal{B}} + H\right).
\end{align}
Therefore, in analogy to the approach extended in \cite{Graham:2009} we find the power spectrum for the inflaton field at the Hubble crossing time $c_{s}k = a H$
\begin{align}
\mathcal{P}^{c=0}_{\phi}\Big|_{c_{s}k= a H} = \frac{{\sqrt{3}HT}}{4\pi\sqrt{\pi} c_{s}}\left(1+Q_{G}\right)^{\frac{1}{2}},
\end{align}
which is exactly the power spectrum obtained in \cite{Motaharfar:2017dxh} for weak dissipation factor ($Q_{G}\ll1$) or $c=0$.

\subsection{Case $c\neq 0$ and $Q_{G}\gg1$}
In the case $c \neq 0$, the inflaton power spectrum can be obtained by solving the coupled inflaton and radiation perturbed field equations

\begin{align}
\delta\phi^{\prime\prime}-(3Q_{G}+2)z^{-1}\delta\phi'
+\delta\phi+3cQ_{G}z^{-2}(\Gamma\dot\phi)^{-1}\delta\rho_r
&=\xi,
\\
\delta\rho_R^{\prime\prime}-(8-c)z^{-1}\delta\rho_R'
+(20-5c)z^{-2}\delta\rho_R+{c^{-2}_{s}}\left(\frac{\delta\rho_{R}}{3}
+(\Gamma\dot\phi)\delta\phi\right)&=0\label{eq13},
\end{align}
where Eq. (\ref{eq13}) is the same as Eq. (\ref{h2}) written in the new time coordinate $z$ and dropping second order perturbations. As we can see, the second equation has been modified by $c_{s}$ in comparison to \cite{Graham:2009}. Therefore, radiation fluctuations may affect the power spectrum in different ways. Analogous to the approach extended in \cite{Graham:2009}, we may find that the power spectrum gets modified with the following factor in the strong regime
\begin{align}
\frac{\mathcal{P}_{\phi}^{c\neq0}}{\mathcal{P}_{\phi}^{c=0}} = \left(\frac{Q_{G}}{Q_{c}}\right)^{3c},
\end{align}
which is different from the results obtained in standard WI scenario since here we have the Galileon dissipation factor $Q_{G}$ instead of dissipation factor $Q$, and $Q_{c}$ is given by
\begin{align}
Q_{c} = \left(\left[\frac{G_{1,3}^{3,1}\Big(\frac{1}{12 c^{2}_{s}}\Big|\begin{smallmatrix}&1-3c/2&\\2-c/2,&0,&5/2\end{smallmatrix}\Big)}{2^{3c} \Gamma_{R}(\frac{3c}{2}) \Gamma_{R}(\frac{3c}{2}+ \frac{5}{2}) \Gamma_{R}(2+c)}\right]^{2} \frac{\Gamma_{R}(3c+\frac{3}{2})}{\Gamma_{R}(\frac{3}{2})}\right)^{-\frac{1}{3c}},
\end{align}
with $G_{1,3}^{3,1}$ and $\Gamma_{R}$ being Meijer-G function and Gamma function, respectively. Therefore, the inflaton power spectrum, valid in both weak and strong regimes with temperature dependent dissipative coefficients, can be written in a less accurate form as follows
\begin{align}
\mathcal{P}^{c\neq0}_{\phi}\Big|_{c_{s}k= a H} = \frac{\sqrt{3}HT}{4\pi\sqrt{\pi}c_{s}} (1+Q_{G})^{\frac{1}{2}} \left(1+ \frac{Q_{G}}{Q_{c}}\right)^{3c},
\end{align}
which is consistent with the old results obtained in previous section for $c=0$ or $Q_{G}\ll1$ and approaches the results obtained for strong dissipation regime for large $Q_{G}\gg1$.

\section{Observable quantities and predictions}\label{Observable quantities and predictions}

\begin{figure}
\begin{center}
\includegraphics[scale=0.85]{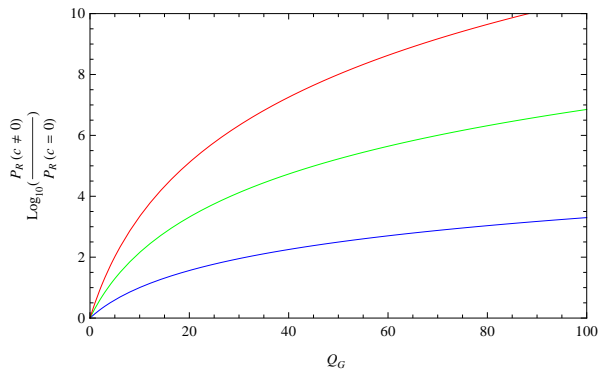}
\caption{An example of variation of the power spectrum with $c\neq0$ normalized by the power spectrum with $c=0$ versus Galileon dissipation factor $Q_{G}$ for $c_{s} =0.9$ where blue, green and red denote $c=1, 2$ and $3$, respectively.}\label{gc}
\end{center}
\end{figure}

\begin{figure}
\includegraphics[scale=0.8]{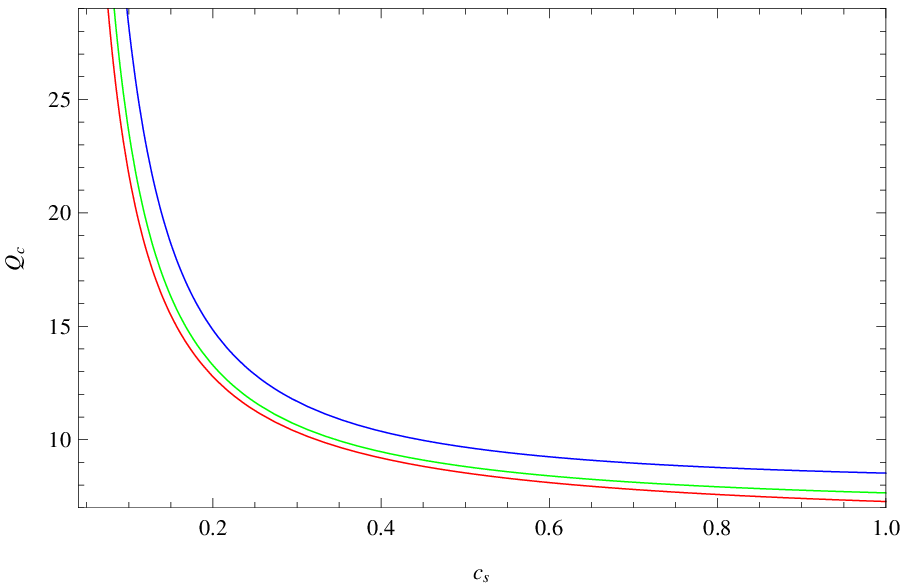} \includegraphics[scale=0.85]{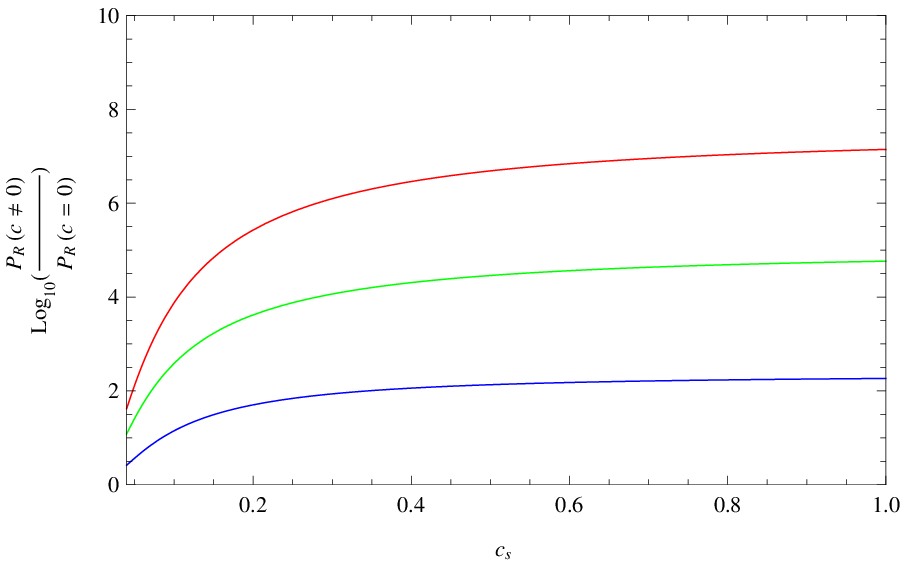}
\caption{Left: Variation of $Q_{c}$ versus $c_{s}$, Right:  An example of variation of power spectrum with $c\neq0$ normalized by the power spectrum with $c=0$ versus $c_{s}$ and for $Q_{G}=40$. In both panels, blue, green and red  denote $c=1, 2$ and $3$, respectively.}
\label{zx}
\end{figure}

In the previous section, we generally derived the inflaton power spectrum for WGI model, but
dissipation processes during inflation imply that both entropy and curvature perturbations must be present in WI scenario. During inflation the energy density of radiation is subdominant such that its thermal fluctuations merely contribute to entropy perturbations; therefore, in such systems with a heat bath, entropy perturbations decay on scales larger than horizon and consequently, one should keep track of curvature perturbations. Primordial cosmological perturbations are typically expressed in terms of curvature perturbation on uniform energy density hypersurfaces denoted by $\mathcal{R}$. The reason behind using this quantity is that it is conserved at large scales in simple models, even beyond linear order perturbation theory. In linear order perturbation theory for the slow-roll single field inflation (warm-G-inflation is dominated by one single canonical field kinetically modified by the Galilean field interaction in over-damped slow-roll regime) the curvature perturbation on the uniform density hypersurface is given by the gauge invariant linear combination $\mathcal{R}= \psi + \frac{H}{\dot \rho}\delta \rho $ with $\psi$ being the spatial metric perturbation and $\delta \phi$ representing perturbations about the homogeneous inflaton field, respectively. For convenience, we choose the spatially flat gauge ($\psi = 0$ which means we have neglected metric perturbations) and accordingly, the curvature perturbation is given by $\mathcal{R}=\frac{H}{\dot\rho}\delta\rho$ which in the slow-roll regime becomes $\mathcal{R} = \frac{H}{\dot\phi} \delta{\phi}$ (in fact, curvature perturbation and inflaton fluctuation are related through this equation) \cite{Zhang:2014}. Therefore, the inflaton power spectrum and curvature power spectrum have the following relation

\begin{align}
\mathcal{P}_{\mathcal{R}} = \frac{H^{2}}{\dot\phi^{2}} \mathcal{P}_{\phi},
\end{align}
and the curvature power spectrum for WGI is given by

\begin{align}\label{rd}
\mathcal{P}^{c\neq0}_{\mathcal{R}}\Big|_{c_{s}k= a H} =  \left(\frac{{\sqrt{3}} H^{3}T}{4\pi\sqrt{\pi} c_{s}\dot\phi^{2}}\sqrt{1+Q_{G}}\right) \left(1+ \frac{Q_{G}}{Q_{c}}\right)^{3c}.
\end{align}
For $G= 0$, we have $c_{s}=1$ and $\mathcal{B}=1$ and the resulting power spectrum reduces to the power spectrum  obtained for standard warm inflation in \cite{Graham:2009}. Figure \ref{gc} shows the logarithmic variation of the power spectrum with $c\neq0$ normalized by power spectrum with $c = 0$ versus $Q_{G}$ for fixed $c_{s}$. As can be seen, the power spectrum is amplified and becomes blue by increasing the value of the Galileon dissipation factor. We have also plotted $Q_{c}$ versus $c_{s}$ in the left panel of figure \ref{zx} for $c= 1,2,3$ where its value is the same as that obtained in \cite{Graham:2009} for $G=0$ or $c_{s}=1$. The striking feature of WGI scenario which distinguishes it from the Standard WI scenario is that although in systems for which dissipation grows during inflation the power spectrum is amplified but such growing mode can be controlled or completely disappeared by decreasing the value of propagating sound speed $c_{s}$ as we have shown in the right panel of figure \ref{zx}. By taking logarithmic derivative of Eq. (\ref{rd}) we obtain the corresponding spectral index

\begin{align}\label{xz}
\nonumber n_{s}-1 &\equiv \frac{d \ln \mathcal{P}_{\mathcal{R}}}{d \ln k} = \frac{\dot{\mathcal{P}}_{\mathcal{R}}}{H\mathcal{P}_{\mathcal{R}}}= -3\epsilon_{H}+ \delta_{T} - \delta_{c_{s}}-2 \delta_{\phi} + \left( \frac{1}{2} \frac{Q_{G}}{1+ Q_{G}} +\frac{3c Q_{G}}{Q_{c}+Q_{G}}\right) \delta_{Q_{G}}\\ &= (n_{s}-1)\big|^{c=0} + \left(\frac{3c Q_{G}}{Q_{c}+Q_{G}}\right) \delta_{Q_{G}},
\end{align}
where $\delta_{Y} \equiv \frac{\dot Y}{H Y}$ and $Y$ being any dynamical parameters of the model and \cite{Motaharfar:2017dxh}

\begin{align}
\delta_{Q_{G}} = \delta_{\Gamma} + \epsilon_{H} - \delta_{\mathcal{B}}, \ \ \ \ \ \ \delta_{\Gamma} = - \frac{\beta}{\mathcal{A}} + \frac{c}{3}\delta_{s}, \ \ \ \ \ \ \ \delta_{T} = \frac{\delta_{s}}{3},
\end{align}
where $s$ is entropy density of the universe and $\delta_{s}$ and $\delta_{\mathcal{B}}$ are very small quantities. As we can observe from Eq. (\ref{xz}) the spectral index is of order $\epsilon$ (here $\epsilon$ indicates first order in perturbations) which means it is scale invariant and  coincides with observation qualitatively. The modified spectral index has an additional term compared to our previous work ($c = 0$) which, in high dissipation regime and when $\epsilon_{H}$ is the dominant component, can positively contribute to the spectral index and consequently, the spectral index can be blue tilted in a high dissipation regime. Furthermore, the corresponding running of the spectral index can be written as

\begin{align}
n_{s}^{\prime}\equiv  \frac{d n_{s}}{d \ln k} = {n_{s}^{\prime}}\big|_{c=0} + \frac{3 c Q_{c}Q_{G}}{\left(Q_{c}+Q_{G}\right)^{2}} \delta^{2}_{Q_{G}} + \frac{3cQ_{G}}{Q_{c}+Q_{G}}\delta_{Q_{G}}\eta_{Q_{G}},
\end{align}
where $\eta_{Q_{G}} \equiv \frac{d \ln \delta_{Q_{G}}}{d \ln k}$ and as can be observed the running is vanishing which means that the size of variations of the spectral index is very small but it has an additional term from the coupling between inflaton and radiation fields in strong dissipation regime which may positively contribute to the running, leading to large positive values of this quantity. In fact, having positive running is one of the interesting features of the WI scenario which distinguishes it from a CI picture \cite{Benetti:2017}. The tensor perturbations do not couple to thermal background and therefore, gravitational waves are merely generated by the quantum fluctuations as in conventional inflation

\begin{align}
\mathcal{P}_{T} = 2M^{-2}_{pl} \left(\frac{H}{2\pi}\right)^{2}.
\end{align}
The corresponding spectral index of gravitational waves is expressed by

\begin{align}\label{tr}
n_{T} = - 2 \epsilon_{H} = - 2 \frac{\epsilon}{\mathcal{A}}.
\end{align}
The tensor-to-scalar ratio may now be written as follows

\begin{align}\label{te}
r = \frac{\mathcal{P}_{T}}{\mathcal{P}_{\mathcal{R}}} = \frac{4c_{s}\epsilon}{\sqrt{3\pi}\mathcal{A}^{2}}\frac{H}{T} \left( 1+Q_{G}\right)^{-\frac12} \left(1+ \frac{Q_{G}}{Q_{c}}\right)^{-3c},
\end{align}
where it is modified by an additional factor compared to our previous work where $r$ could assume smaller values in the high dissipation regime. Considering Eqs. (\ref{tr}, \ref{te}), we may derive the consistency relation as follows

\begin{align}
r =- \frac{2c_{s}}{\sqrt{3\pi}\mathcal{A}}\frac{H}{T} \left( 1+Q_{G}\right)^{-\frac12} \left(1+ \frac{Q_{G}}{Q_{c}}\right)^{-3c} {n_{T}},
\end{align}
which is not a constant relation, in contrast to cold G-inflation. The relation for $\frac{T}{H}$ also reads as

\begin{align}
\frac{T}{H} &= \left(\frac{3\sqrt{3}\mathcal{B} Q_{G}}{16\pi\sqrt{\pi} \mathcal{P}_{\mathcal{R}}C_{R}c_{s}}\right)^{\frac{1}{3}} \left(1+Q_{G}\right)^{\frac16}\left(1+ \frac{Q_{G}}{Q_{c}}\right)^{c}\label{tf}.
\end{align}
Hence, the condition ($T>H$) for WI scenario to occur can be obtained by $Q_{G}> C_{R}\mathcal{P}_{\mathcal{R}}$. Taking $g_{\star}$ of order $10^{2}$ and $\mathcal{P}_{\mathcal{R}}$ of order $10^{-9}$, we deduce that a small amount of Galileon dissipation results in warm inflation. Furthermore, thanks to the growing mode function,  $\frac{T}{H}$ may obtain larger value in the strong regime. One may evaluate the variations of the inflaton field for observable scales with $\Delta N \simeq 4$ corresponding to multipoles $2<l<100$ at the horizon crossing time as follows

\begin{align}
\frac{\Delta\phi}{M_{pl}} &= \frac{\dot\phi\Delta N}{M_{pl}H} \simeq \left(12\pi\right)^{\frac14}c_{s}^{-\frac12}\left( 1+Q_{G}\right)^{\frac14} \left(1+ \frac{Q_{G}}{Q_{c}}\right)^{\frac{3c}{2}} \left(\frac{T}{H}\right)^{\frac12} r^{\frac{1}{2}}\label{tg}.
\end{align}
In fact, such a parameter accommodates the inflaton field in two classes; first, large field models where $\Delta\phi\gg M_{pl}$ and second, small field models where $\Delta\phi\ll M_{pl}$. As it is clear, the model can obtain large field excursions in strong Galileon dissipation factor even if the tensor-to-scalar ratio is very small where the over-large amplitude of the inflaton field in the CI picture may now be solved more easily due to the presence of a growing mode function. Such an elegant feature is also present in the non-G-inflation limit.

\section{Warm Higgs G-inflation }\label{Warm Higgs G-inflation}

In a Warm Higgs G-inflation (WHGI) construction we need to add a Galileon interaction term to the SM Higgs Lagrangian in a WI scenario, therefore

\begin{align}
S_{WHGI} = \int d^{4}x \sqrt{-g} \left[\frac{M^{2}_{pl}}{2}R - |D_{\mu}\mathcal{H}|^{2} - \lambda (|\mathcal{H}|^{2}-\nu^{2})^{2}-  \frac{2\mathcal{H}^{\dag}}{M^{4}} D^{\mu}D_{\mu}\mathcal{H} |D_{\mu}\mathcal{H}|^{2} + \mathcal{L}_{R}+ \mathcal{L}_{int}\right],
\end{align}
where $D_{\mu}$ is the covariant derivative with respect to the SM gauge symmetry, $\mathcal{H}$ is the SM Higgs boson, $\nu$ is the vacuum expectation value ($vev$) of the SM Higgs and $\lambda$  is the self coupling constant. Since we would like to have a chaotic inflation-like dynamics of the Higgs boson, we consider the case where its neutral component $\phi = \sqrt{2}|\mathcal{H}_{0}|$ is very large compared to the electroweak vacuum expectation value $\nu = 246$ Gev ($\phi\gg\nu$). Therefore, we should only consider a simpler action as follows

\begin{align}
S_{WHGI} = \int d^{4}x \sqrt{-g} \left[\frac{M^{2}_{pl}}{2}R - X - \frac{\phi X}{M^{4}} \Box \phi + \frac{\lambda}{4}\phi^{4}+ \mathcal{L}_{R}+ \mathcal{L}_{int}\right],
\end{align}
where $M$ has the dimension of mass ($M$>0). It deserves to be noted that the gauge fields which couple to the neutral component $\phi$ receive heavy mass from the field value of the Higgs boson and hence we can neglect the effect of gauge fields when we consider the inflationary trajectory. Thus, this setup corresponds to the case

\begin{align}
G(\phi, X) = - \frac{\phi X}{M^{4}},~~ V(\phi) = \frac{\lambda}{4}\phi^{4}\label{pe},
\end{align}
where in order to further analyze the predictions of the model we consider more general form of the Galileon interaction term inspired from quantum field theory where it may have a power-law form for both $\phi$ and $X$, therefore

\begin{align}\label{pe2}
G(\phi, X) = - \frac{\phi^{2p+1}X^{q}}{M^{4q+2p}},
\end{align}
where $p$ and $q$ are positive. To complete our setup we should determine how Galileon scalar field dissipates to radiation during inflation. To this end, we consider a linear temperature dissipation coefficient as follows

\begin{align}\label{wq}
\Gamma(\phi, T) = C_{T} T,
\end{align}
where $C_{T}$ is connected to the dissipative microphysics dynamics. The most elegant feature of this form of dissipative coefficient may be realized in the spirit of ``Littel Higgs'' models when the inflaton is a pseudo-Nambo Goldstone boson of a broken gauge symmetry, its $T=0$ potential being protected against large radiative corrections by symmetry while still having enough interactions to allow thermalization of light degrees of freedom and resulting in enough dissipation even if the mediators are so light; an example would be fermions directly coupled to the inflaton \cite{Bastero-Gil:2016qru}.

Since we are interested in understanding the role played by the Galileon interaction term in the dynamics of WGHI, we solve the model in the G-dominant regime which means $|\delta_{GX}|\gg|\delta_{X}|$, thus utilizing Eqs. (\ref{t2}), the Higgs self interaction (\ref{pe}) and generalized Galileon interaction term (\ref{pe2}), we obtain velocity of the inflaton field in terms of $Q_{G}$ and $\phi$ given by

\begin{align}\label{re1}
\dot\phi = -  M^{2}_{pl} \zeta_{1}^{\frac{1}{2q}} \left(Q_{G}+\frac{1}{2q}\right)^{-\frac{1}{2q}} \left(\frac{\phi}{M_{pl}}\right)^{-\frac{(p+1)}{q}},
\end{align}
where $\zeta_{1}= \frac{2^{q}}{3q^{2}} \left(\frac{M}{M_{pl}}\right)^{4q+2p}$ and we have considered minus signature for the field velocity in order to obtain positive $\delta_{GX}$. Using Eqs.(\ref{re1}) and (\ref{t1}), we obtain a relation between the inflaton field and Galileon dissipation factor as follows

\begin{align}\label{az3}
Q_{G}^{3}\left(Q_{G}+\frac{1}{2q}\right)^{\frac{5-6q}{2q}} = \zeta_{2} \left(\frac{\phi}{M_{pl}}\right)^{-\frac{11q+5p+5}{q}},
\end{align}
with $\zeta_{2} =\frac{\sqrt{3}C_{T}^{4}}{4 C_{R}\lambda^{\frac72}}\zeta_{1}^{\frac{5}{2q}} $. Now, using Eqs. (\ref{re1}) and (\ref{az3}) we can write all parameters of the model as a function of $Q_{G}$, therefore, for later convenience we write $\mathcal{B}$ as

\begin{align}
\mathcal{B} = \zeta_{3}  Q_{G}^{-3a} \left(Q_{G}+\frac{1}{2q}\right)^{\frac{2q(3a-1)-5a+1}{2q}},
\end{align}
where $a = \frac{q+p+1}{11q+5p+5}$ and $\zeta_{3} =\sqrt{\frac{4\lambda}{3}}\zeta_{1}^{-\frac{1}{2q}} \zeta_{2}^{a} $. One may find the evolution of the inflaton field with respect to e-folding number using Eq. ($\ref{az}$) in terms of $\phi$ and $Q_{G}$

\begin{align}\label{az2}
\frac{d \ln \frac{\phi}{M_{pl}}}{d N} = - \sqrt{\frac{12}{\lambda}} \zeta_{1}^{\frac{1}{2q}} \left(Q_{G}+\frac{1}{2q}\right)^{-\frac{1}{2q}} \left(\frac{\phi}{M_{pl}}\right)^{-\frac{3q+p+1}{q}},
\end{align}
where a minus sign means that we evaluate the number of e-folding from the end of inflation. Then, taking the derivative with respect to the e-folding number from Eq. (\ref{az3}) we can obtain the evolution of the Galileon dissipation factor with respect to $N_{e}$ as follows

\begin{align}\label{sx}
\frac{dQ_{G}}{dN_{e}} = \zeta_{4} \left(22q+10p+10\right) \frac{Q_{G}^{3b+1}\left(Q_{G}+\frac{1}{2q}\right)^{-\frac{2q(3b-1)-5b+1}{2q}}}{{5}Q_{G}+ {3}},
\end{align}
where $b = \frac{3q+p+1}{11q+5p+5}$ and $\zeta_{4} = 2\sqrt{\frac{3}{\lambda}} \zeta_{1}^{\frac{1}{2q}} \zeta_{2}^{-b}$. Integrating Eq. (\ref{sx}) we obtain a relation between $N_{e}$ and $Q_{G}$ given by

\begin{align}\label{ob1}
\zeta_{4}N_{e} = f(Q^{end}_{G})-f(Q^{hc}_{G}),
\end{align}
where $Q_{G}^{hc}$ and $Q_{G}^{end}$ denote the value of the Galileon dissipation factor at the Hubble crossing time and at the end of inflation, respectively, and $f(Q_{G})$ is given by

\begin{align}
\nonumber f(Q_{G}) &=-\frac{q Q_{G}^{-3 b} \left(\frac{1}{2 q}+Q_{G}\right)^{\frac{6 b q-5 b+1}{2 q}} \left(5-\frac{\, _2F_1\left(1,\frac{1-5 b}{2 q};1-3 b;-2 q Q_{G}\right)}{b}\right)}{(5 b-1) (5 p+11 q+5)}.
\end{align}
As is clear from Eq. (\ref{ob1}), we have a relation for the number of e-folding without consideration of being in a weak or strong dissipation regime since as we know that the dissipation factor evolves during inflation in the WI scenario. Hence, it may start with a small value and becomes larger until the end of inflation \cite{\iffalse WI PRL\fi Bastero-Gil:2016qru}. Furthermore, the evolution of other dynamical parameters of the model may be written as a function of $Q_{G}$ as

\begin{align}
\frac{T}{M_{pl}} &= \left(\frac{\sqrt{3\lambda}}{2C_{R}} \zeta^{\frac{1}{2q}}\zeta_{2}^{4d}\right)^{\frac14} Q_{G}^{\frac{-3d+1}{4}} \left(Q_{G}+ \frac{1}{2q}\right)^{\frac{2q(3d-1)-5d-1}{8q}} \label{tt1},\\
\frac{\rho_{R}}{V}& = {{\zeta}_{4}} Q_{G}^{3b+1} \left(Q_{G}+ \frac{1}{2q}\right)^{\frac{-2q(3b+1)+5b-1}{2q}}\label{tt2},\\
\frac{T}{H} &= \left(\frac{3\sqrt{3}\zeta_{3}}{16\pi\sqrt{\pi} \mathcal{P}_{\mathcal{R}}C_{R}c_{s}}\right)^{\frac{1}{3}}  Q_{G}^{-a+1} \left(1+Q_{G}\right)^{\frac16}\left(1+ \frac{Q_{G}}{Q_{c}}\right)\left(Q_{G}+\frac{1}{2q}\right)^{\frac{2q(3a-1)-5a+1}{6q}}\label{t3},
\end{align}
where $d= \frac{q-p-1}{11q+5p+5}$. Also, the observable quantities of the models $\mathcal{P}_{R}$, $n_{s}$ and $r$ may be written as a function of $Q_{G}$
\begin{align}
\mathcal{P}_{\mathcal{R}} &= \left(\frac{C_{T}^{3}}{48\pi\sqrt{3\pi}\zeta_{3}^ {2}C_{R}c_{s}}\right)Q_{G}^{6a-2}\left({1+Q_{G}}\right)^{\frac12}\left(1+ \frac{Q_{G}}{Q_{c}}\right)^{3}\left(Q_{G}+\frac{1}{2q}\right)^{\frac{-2q(3a-1)+5a-1}{q}}\label{ob4},\\
\nonumber n_{s}-1  &= \frac{d \ln \mathcal{P}_{\mathcal{R}}}{d \ln k} = \nonumber  \frac{d \ln \mathcal{P}_{\mathcal{R}}}{d Q_{G}} \frac{d Q_{G}}{d N} =\zeta_{4}\left(22q+10p+10\right) \left(\frac{Q_{G}^{3b+1}\left(Q_{G}+\frac{1}{2q}\right)^{-\frac{2q(3b-1)-5b+1}{2q}}}{{5}Q_{G}+ {3}}\right)\\ & \times\left[\frac{6a-2}{Q_{G}}+\frac{1}{2(1+Q_{G})}+\frac{3}{Q_{c}+Q_{G}}+ \frac{-2q(3a-1)+5a-1}{q(Q_{G}+\frac{1}{2q})}\right]\label{ob2}, \\
r& = \frac{\mathcal{P}_{T}}{\mathcal{P}_{\mathcal{R}}} = \frac{\lambda\zeta_{2}^{f}}{24\pi^{2}\mathcal{P}_{\mathcal{R}}} (Q_{G})^{-3f}\left(Q_{G}+\frac{1}{2q}\right)^{\frac{(6q-5)f}{2q}},
\end{align}
with $f = \frac{4q}{11q+5p+5}$. One may also obtain the field excursion given by

\begin{align}
\left|\frac{\Delta\phi}{M_{pl}}\right|\simeq \left|\frac{\dot\phi \Delta N}{M_{pl}H}\right| \simeq \zeta_{5} (Q_{G})^{3c} \left(Q_{G}+\frac{1}{2q}\right)^{\frac{(5-6q)c-1}{2q}},
\end{align}
 with $\zeta_{5} = 8 \sqrt{\frac{3}{\lambda}} \zeta_{1}^{\frac{1}{2q}} \zeta_{2}^{-c}$,~ $c = \frac{2q+p+1}{11q+5p+5} $. Furthermore, the two conditions $|\delta_{GX}|\gg |\delta_{X}|$ (G-dominant regime) and $|\delta_{G\phi}|\ll |\delta_{X}|$ can be combined as $|\delta_{G\phi}|\ll |\delta_{GX}|$. This means that the generalized Galileon interaction term is kinetically dominated, therefore, we also obtain the following ratio as a function of $Q_{G}$

\begin{align}
\left|\frac{\delta_{G\phi}}{\delta_{GX}}\right| = \frac{2p+1}{2q} \zeta_{4}  (Q_{G})^{3b}\left(Q_{G}+\frac{1}{2q}\right)^{\frac{(5-6q)b-1}{2q}},
\end{align}
which should be small. Finally to obtain the value of $Q_{G}$ at the end of inflation we need to solve equation $\epsilon_{H}=1$ which can be written as

\begin{align}\label{ob3}
 \frac{1}{2} \zeta_{4}^{-1} {(Q_{G}^{end})}^{-3b} \left(Q^{end}_{G}+\frac{1}{2q}\right)^{\frac{(6q-5)b+1}{2q}} =1.
\end{align}
Before closing the section, it should be mentioned that to evaluate all the parameters of a particular model (for fixed $p$ and $q$) we require to fix $Q_{G}, \lambda, C_{T}$ and $M$ and therefore we attempt to constrain these parameters with observational data.

\section{Observable quantities and constraints} \label{Observable quantities and constraints}

In this section, we will present the methodology for analyzing the parameter dependence of the WGI model on inflationary observable quantities and evolution of homogeneous dynamical quantities of the model during inflation. In particular, we will consider the scalar amplitude $A_{s}$ and its spectral index $n_{s}$ as observable quantities which are functions of three constant parameters; (i) $\lambda$, the coupling constant for the Higgs self-interaction potential, (ii) $C_{T}$, the proportionality constant for the dissipative ratio, and (iii) $M$, the coupling constant determining the strength of the Galileon interaction term, although other than these constant parameters they depend on the dissipative ratio $Q_{G}$ which should be calculated at the Hubble crossing time. Also, these observable quantities depend on the parameters $p$ and $q$ (or correspondingly $c_{s}$) which determines a particular WGI scenario as we have shown in the previous section.

The reproduction of cosmological observational data by an inflationary model not only implies the validity of that particular model but also provides the opportunity to constrain the parameters of the model in a more physical manner. Therefore, more precise observational data regarding Cosmic Microwave Background (CMB) map not only helps to distinguish inflationary models (or ruling out invalid ones) but also narrows the range of parameters involved. In this respect, several collaborations have tried to obtain new observational constraints on observable parameters using the recently released Planck 2015 data. As a matter of fact, joining Planck likelihood with $TT$, $TE$ and $EE$ polarization modes plus BAO likelihood give $n_{s}= 0.9656\pm 0.00825$, $\alpha_{s} = -0.00885\pm 0.01505$, $r<0.1504$ and the normalizing scalar perturbation amplitude $A_{s} = 2.24 \times 10^{-9}$ at $95\%$ confidential level \cite{Ade:2015xua}.

To have a viable inflationary model requires a sufficient number of e-folding in order to solve the flatness problem. The number of e-folding $N_{e}$ depends on the energy scale of inflation and may take different values. In fact, inflation may take place at the electroweak energy scale ranging from $10^{16} Gev$ at the highest to $1 Mev$ at the lowest. Consequently, we have to make sure that the parameters we choose for the power spectrum and spectral index give a sufficient number of e-folding, i.e.  ($35< N_{e}<65$ \cite{Visinelli:2016}). To this end, we consider $N_{e} = 50$ throughout this paper.

Therefore, to obtain constraints on the parameters of the model we consider the WGI scenario with a central value for the spectral index $n_{s} = 0.9656$ and scalar amplitude normalization $\mathcal{P}_{\mathcal{R}} = A_{s}= 2.24 \times 10^{-9}$ for $N_{e}=50$ at the Hubble crossing time. Then,  Eqs. (\ref{ob1}, \ref{ob4}, \ref{ob2}) and (\ref{ob3}) may be solved to obtain the unknown parameters of the model. To this end, we first numerically solve Eqs. (\ref{ob1}, \ref{ob2} and \ref{ob3}) for fixed values of $N_{e}$, $n_{s}$, $p$ and $q$ to obtain $Q_{Q}^{hc}, Q_{G}^{end}$ and $\zeta_{4}$. Next, using Eq. (\ref{ob4}) and the relation for $\zeta_{4}$ for a fixed value of the scalar amplitude we have two equations and three unknown parameters whereby two parameters may be obtained as a function of the third one. This means that we may obtain the evolution of the parameters versus, for instance, $C_{T}$ which will be discussed in the next section. Therefore, to constrain the parameters of the model we need to fix one unknown parameter in order to obtain the other two. As we mentioned earlier, the crucial condition for warm inflation to happen is $T> H$. Hence, we can fix one parameter by imposing this condition and derive the other two. On the other hand from quantum field theory point of view, the self coupling of the self-interaction Higgs potential may be as large as $ 0.13$. One may then obtain the lower and upper limit on the parameters of the models utilizing these two conditions.

Up to now we have discussed how to fix the parameters of the model in such a way that the model becomes consistent with Planck data at the Hubble crossing time. To obtain the evolution of  homogeneous dynamical quantities, $\frac{\phi}{M_{pl}}, \frac{T}{M_{pl}}, \frac{T}{H}$ and $\frac{\rho_{R}}{V}$, and also observable parameters $\mathcal{P}_{\mathcal{R}}$ and $n_{s}$ we use Eq. (\ref{ob1}). In fact, Eq. (\ref{ob1}) implies that decreasing the number of e-folding means that the system is evolving all the way to $N_{e}=0$ which corresponds to the end of inflation. Hence, we vary the number of e-folding from $0$ to $50$ and obtain the corresponding value of $Q_{G}$ by numerically solving Eq. (\ref{ob1}). One may then insert the obtained values of $Q_{G}$ in Eqs. (\ref{az3}, \ref{tt1}, \ref{tt2}, \ref{t3}, \ref{ob4}) and (\ref{ob2}) and find the evolution of homogeneous and inhomogeneous parameters of the model during inflation.

\section{Results and Discussion } \label{Results and Discussion}

\begin{figure}
\begin{center}
\ \ \ \includegraphics[scale=0.9]{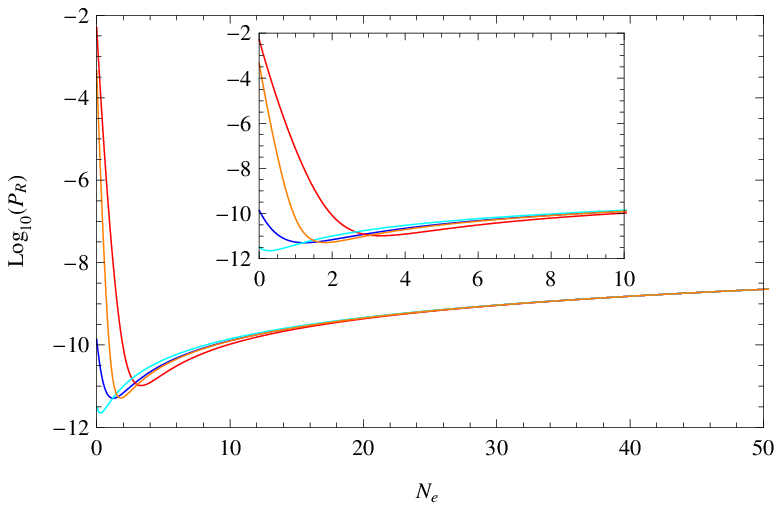}\includegraphics[scale=0.9]{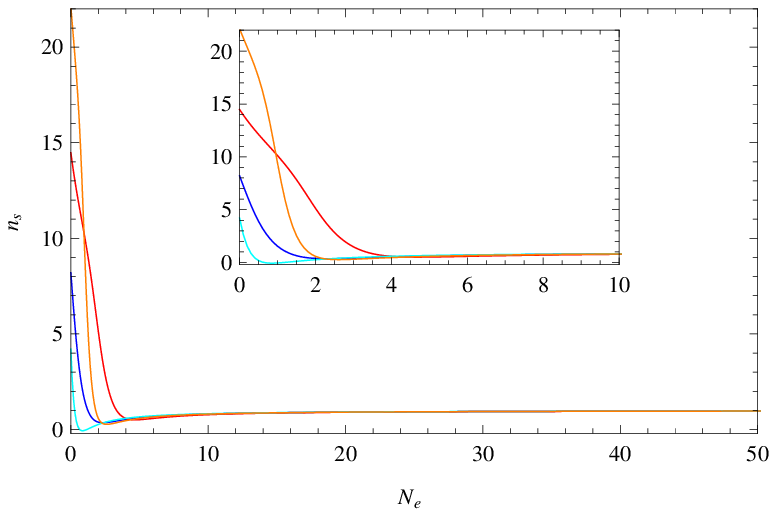}
\caption{The evolution of $\mathcal{P}_{\mathcal{R}}$ and $n_{s}$ versus the number of e-folding $N_{e}$ where blue, cyan, red and orange represent $(p, q) =  (0,1), (0, 2), (1, 1)$ and $(1, 2)$, respectively. Both panels have been plotted for $\lambda = 0.13$, $g_{\star} = 100$ and for values of $C_{T}$ and $M$ presented in Table \ref{l1}.}\label{ff1}
\end{center}
\end{figure}
\begin{figure}
\begin{center}
\includegraphics[scale=0.9]{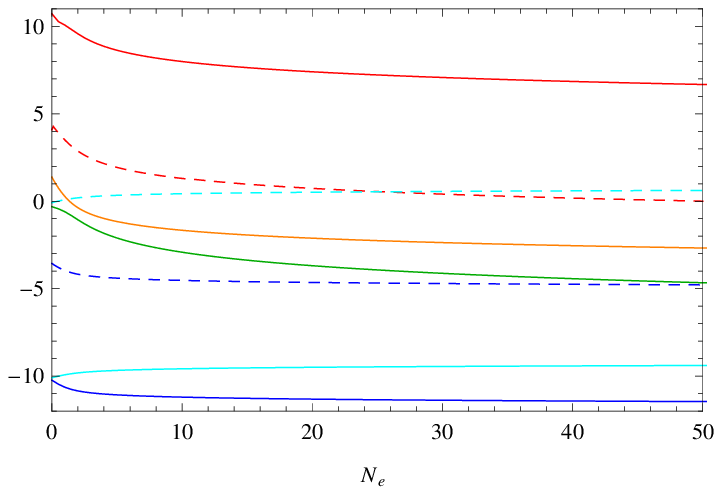} \ \  \ \ \ \ \includegraphics[scale=0.9]{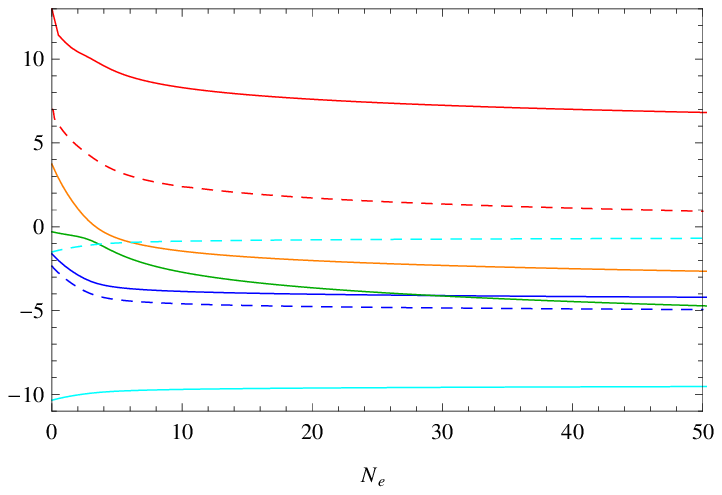} \\
 \includegraphics[scale=0.9]{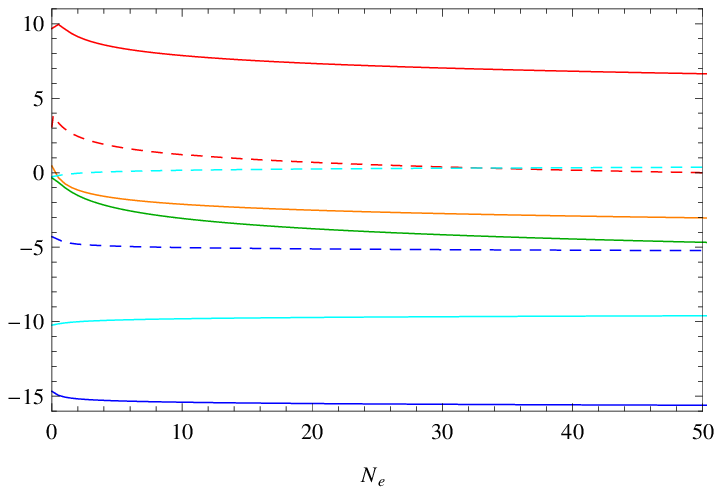} \ \ \ \ \ \ \includegraphics[scale=0.9]{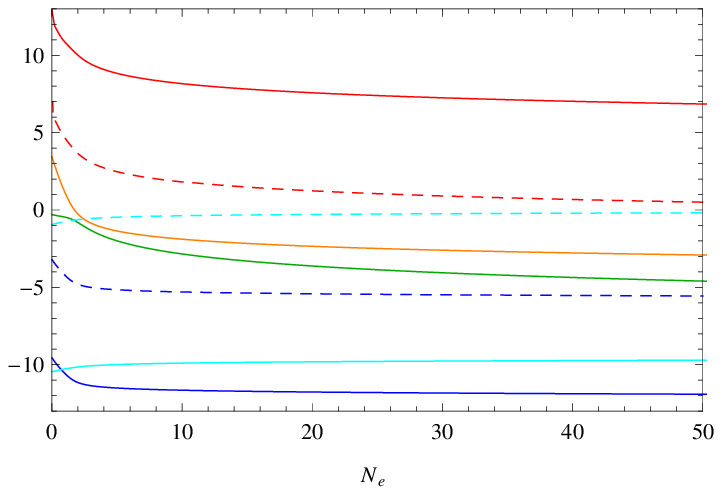}
\end{center}
\caption{Examples of the evolution of dynamical quantities versus the number of e-folding $N_{e}$ during inflation where red, orange, green, cyan and blue represent the logarithmic values of $\frac{T}{H}$, $Q_{G}$, $\frac{\rho_{R}}{V}$, $\frac{\phi}{M_{pl}}$ and $\frac{T}{M_{pl}}$. The left top and down and the right top and down  panels have been plotted for $(p, q) = (0,1), (0, 2), (1, 1)$ and (1, 2), respectively. Solid curves are the result of using the values of parameters in Table \ref{l1} and dashed curves for the values of parameters in Table \ref{l2}.}\label{ff2}
\end{figure}

In this section, we present and discuss the results of the numerical analysis carried out by following the procedure outlined in the  previous section. Let us start by discussing figure \ref{ff1} which shows the evolution of power spectrum and spectral index versus number of e-folding and for four different Galileon interaction terms in which we have fixed the parameters of the model for the values showed in Table \ref{l1} in order to be consistent with Planck data producing $N_{e}=50$. The evolution of the power spectrum and its spectral index show striking features which distinguish WGI from its CI counterpart; 1- They are non-monotonic as in WI scenario and, in fact,  decreasing for small dissipation factor and increasing as approaching $1$. Therefore, the model is in good agreement with observational data at large scales but during inflation  with increasing values of dissipation factor, the power spectrum will be amplified even by several orders of magnitude (this is  consistent with the result obtained for standard WI in \cite{Bastero-Gil:2016qru}).  2-  As  is obvious from Figure \ref{ff1}, the Galileon dissipation factor at the end of inflation may assume smaller values by decreasing the value of the speed of sound $c_{s}$ and for fixed $p$. As a result, the power spectrum won't be amplified much. In fact, the speed of sound parameter can control the amplification of the power spectrum at the end of inflation.

\begin{figure}
\begin{center}
\includegraphics[scale=0.9]{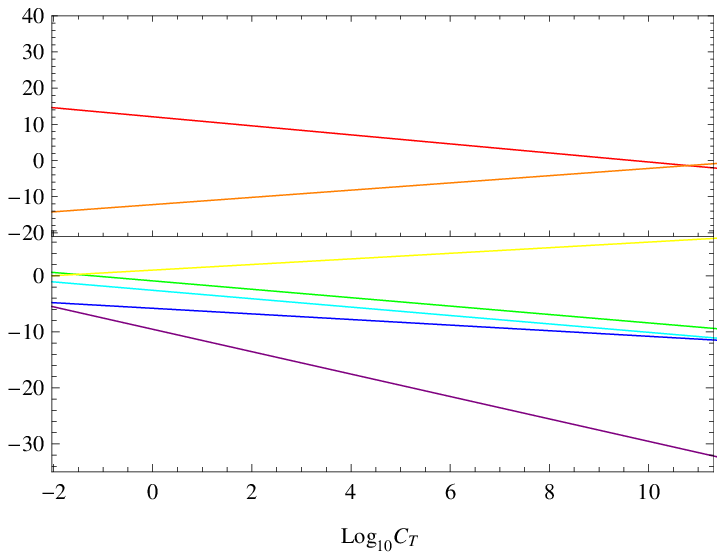} \ \ \ \ \ \  \includegraphics[scale=0.9]{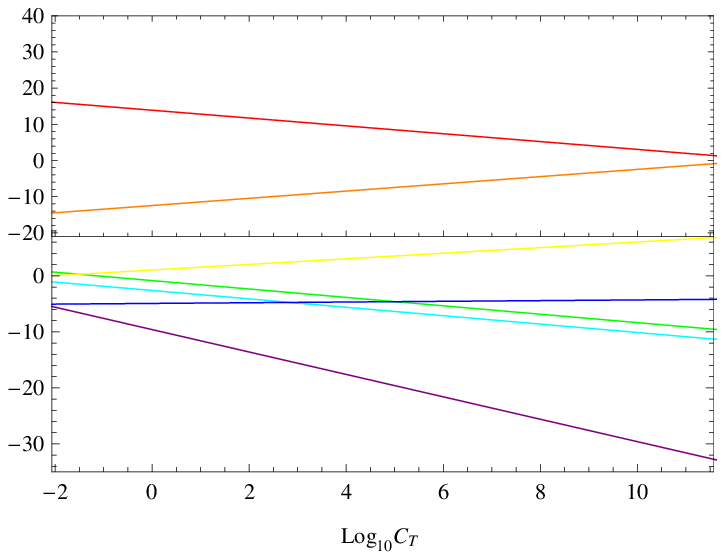} \\
 \includegraphics[scale=0.9]{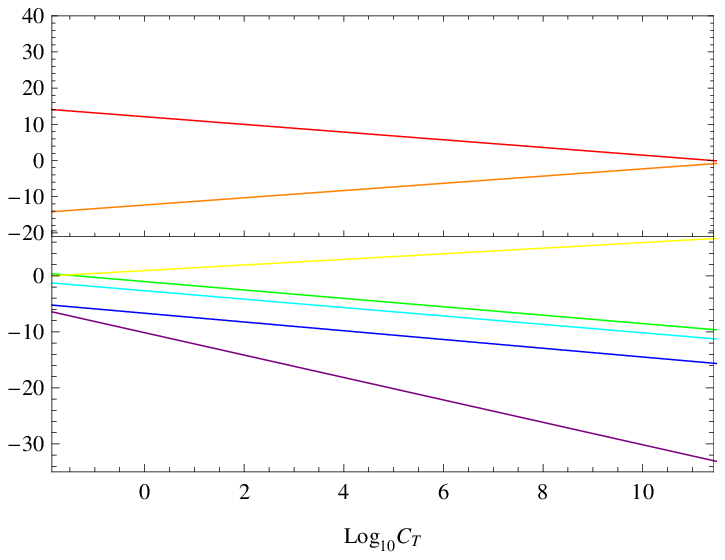} \ \ \ \ \ \ \includegraphics[scale=0.9]{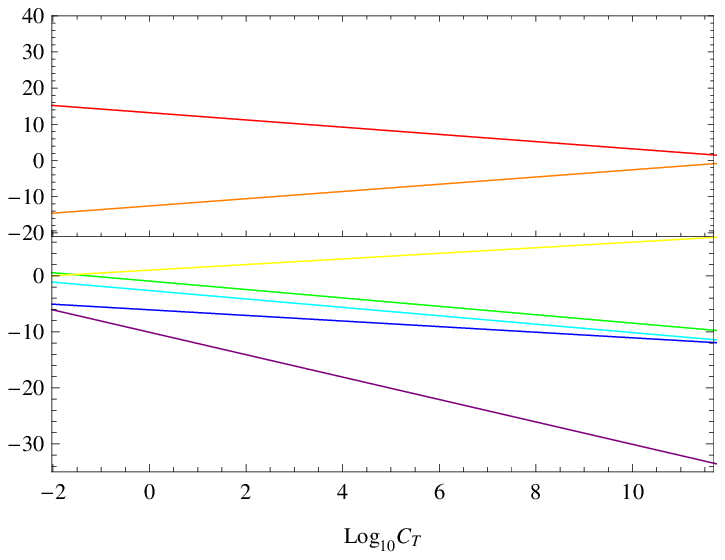}
\end{center}
\caption{The log-log plot for the evolution of the parameters of the model versus $C_{T}$ in which red, orange, yellow, green, cyan, blue and purple represent $ M, \lambda, \frac{T}{H}, \frac{\phi}{M_{pl}}, \frac{\Delta\phi}{M_{pl}}, \frac{T}{M_{pl}}$ and $r$, as well as the left top and down and the right top and down  panels have been plotted for $(p, q) = (0, 1) , (0, 2), (1, 1)$ and (1, 2), respectively.}\label{ff3}
\end{figure}
\begin{figure}
\begin{center}
\includegraphics[scale=0.89]{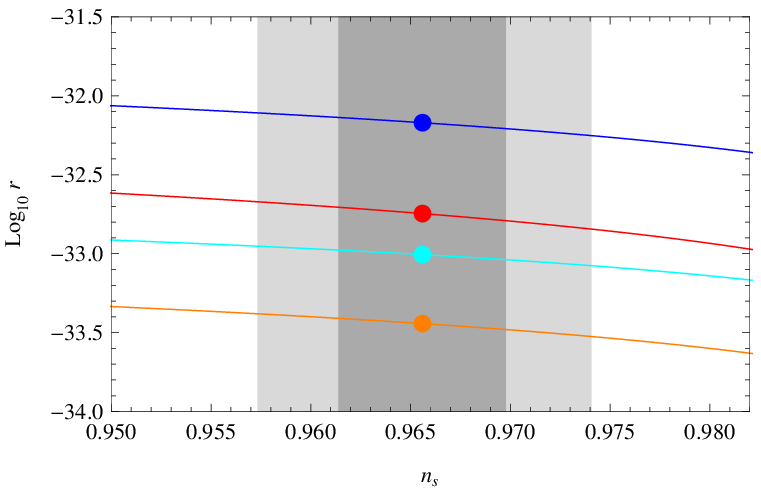} \ \ \ \ \ \ \includegraphics[scale=0.89]{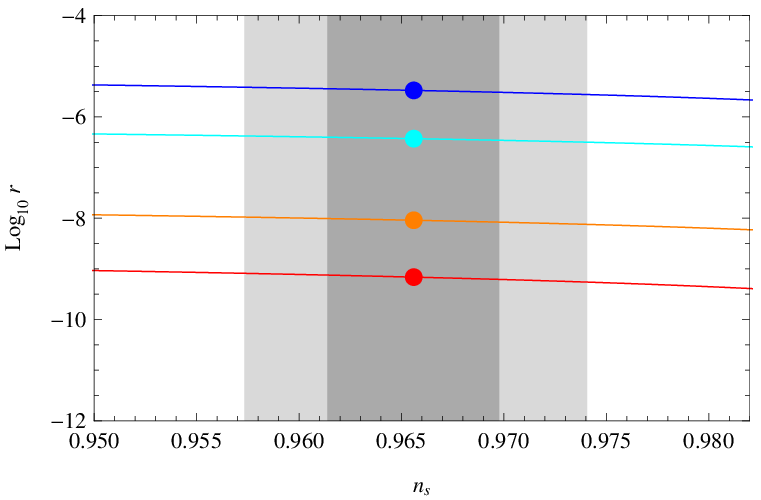}
\end{center}
\caption{The $(r, n_{s})$ plane where  blue, cyan, red and orange represent theoretical predictions for $(p, q) =  (0,1), (0, 2), (1, 1)$ and $(1, 2)$ with variation of $N_{e}$ along the curves where dots denote $N_{e}=50$  as well as dark and light shades representing the range of tilt spectral index for $1 \sigma$ and $2 \sigma$ of Planck likelihood+ TTTEEE+ BAO. Also, we have plotted the Left panel for the values given in Table \ref{l1} and the Right panel for the values given in Table \ref{l2}.}\label{ff4}
\end{figure}

\begin{table}\begin{center}
\begin{tabular}{ |l|l|l|l|l|}
\hline
$ \ \ \ (p,q)   $& $\ \ \ \ \ \ (0,1)$& $\ \ \ \ \ \ (1,1)$& $\ \ \ \ \ \ (0,2)$ &$\ \ \ \ \ \ (1,2)$\\
\hline
$\ \ \ \ \ M$& $ \ \ \ \ 0.0087$ &$ \ \ \ \ \ 22.9866$&$ \ \ \ \ \ 0.9169$&$ \ \ \ \ \ 34.2465$ \\
\hline
$\ \ \ \ \ C_{T}$& $ \ \ 2.06 \times 10^{11}$ &$ \ \ 3.74 \times 10^{11}$&$ \ \ 2.67 \times 10^{11}$&$ \ \ \ 4.8 \times 10^{11}$ \\
\hline
$\ \ \  T/M_{pl}$& $ \ \ 3.5 \times 10^{-12}$ &$ \ \ \ 6.2 \times 10^{-5}$&$ \ \ 2.47\times 10^{-16}$&$ \ \ 1.23 \times 10^{-12}$ \\
\hline
$\ \ \  \phi/M_{pl}$ & $4.075 \times 10^{-10}$ &$ \ 2.92 \times 10^{-10}$&$\ 2.51 \times 10^{-10}$&$\ 1.95 \times 10^{-10}$ \\
\hline
$\ \ \ \ T/H$ & $ \ \ 4.71 \times 10^{6}$ &$ \ \ \ 6.58 \times 10^{6}$&$ \ \ \ 4.4 \times 10^{6}$&$ \ \ \ 7.09 \times 10^{6}$ \\
\hline
 $\ \ \ \ \ Q_{G}$& $ \ \ \ 0.002119$ &$ \ \ \ \ 0.00223$&$ \ \ \ \ 0.000937$&$ \ \ \ \ 0.00123$ \\
\hline
$\ \ \ \ \rho_{R}/V$ & $ \ \ 2.18 \times 10^{-5}$ &$ \ \ 1.9 \times 10^{-5}$&$ \ \ 2.14 \times 10^{-5}$&$ \ \ 2.53 \times 10^{-5}$ \\
\hline
$\ \ \Delta\phi/M_{pl}$& $\ 8.44\times 10^{-12}$ &$ \ 5.05 \times 10^{-12}$&$ \ 5.78\times 10^{-12}$&$ \ 4.05 \times 10^{-12}$ \\
\hline
$ \ |{\delta_{G\phi}}/{\delta_{GX}}| $  & $ \ \ \ \ 0.00258$ &$ \ \ \ \ 0.00647$&$ \ \ \ \ 0.00143$&$ \ \ \ \ 0.00387$ \\
\hline
$\ \ \ \ \ \  r$ & $ \ 6.75\times 10^{-33}$ &$ \ \ 1.79\times 10^{-33}$&$ \ \ 9.86 \times 10^{-34}$&$ \ \ 3.6 \times 10^{-34}$ \\
\hline
$\ \ \ \ \  r/n_{t}$ & $ \ -1.3\times 10^{-30}$ &$ \ -4.15\times 10^{-31}$&$ \ -1.71 \times 10^{-31}$&$ \ -6.97 \times 10^{-32}$ \\
\hline
$\ \ \ \ \  n^{\prime}_{s}$& $ \ \  -0.00207$ &$ \ \ \ -0.00333$&$ \ \ \ -0.0014$&$ \ \ \ -0.00232$ \\
\hline
\end{tabular}
\caption{The values of the parameters of the models for $\mathcal{P}_{\mathcal{R}} = 2.24 \times 10^{-9}$, $n_{s} = 0.9656$, $N_{e} = 50$, $\lambda = 0.13$, $g_{\star} = 100$ and four different $(p,q)$.}\label{l1}
\end{center}\end{table}
\begin{table}\begin{center}
\begin{tabular}{ |l|l|l|l|l|}
\hline
$ \ \ \ (p,q)   $& $\ \ \ \ \ \ (0,1)$& $\ \ \ \ \  \ (1,1)$& $\ \ \ \ \ \  (0,2)$ &$\ \ \ \ \ \ (1,2)$\\
\hline
$\ \ T/M_{pl}<$& $ \  1.65 \times 10^{-5}$ &$ \ 1.14 \times 10^{-5}$&$ \ 5.95\times 10^{-6}$&$ \ 2.76 \times 10^{-6}$ \\
\hline
$\ \ \phi/M_{pl}<$ & $\ \ \ \ \ 4.173$ &$\ \ \ \ \ 0.204$&$\ \ \ \ \ 2.331$&$\ \ \ \ \ 0.657$ \\
\hline
$\Delta\phi/M_{pl}<$& $\ \ \ \ \ 0.0864$ &$\ \ \ \ \ 0.00352$&$\ \ \ \ \ 0.0535$&$\ \ \ \ \ 0.013$ \\
\hline
$\ \ \ \  r<$ & $ \ 3.34\times 10^{-6}$ &$ \ 6.87\times 10^{-10}$&$ \ 3.724 \times 10^{-7}$&$ \ 9.12 \times 10^{-9}$ \\
\hline
$\ \ \ \  r/n_{t}<$ & $  -6.45\times 10^{-4}$ &$  -1.59\times 10^{-7}$&$  -6.48 \times 10^{-5}$&$ -1.76 \times 10^{-6}$ \\
\hline
$\ \ \ \  \lambda>$& $ \ 5.84 \times 10^{-15}$ &$ \ 2.1 \times 10^{-13}$&$ \ 6.69 \times 10^{-15}$&$ \ 2.58 \times 10^{-14} $ \\
\hline
$\ \ \ \ C_{T}>$& $ \ \ \ \ \ 0.00926$ &$ \ \ \ \ \  0.605$&$ \ \ \ \ \ 0.0137$&$ \ \ \ \ \ 0.0955$ \\
\hline
$ \ \ \ \ M $<& $\ 4.23\times 10^{14}$ &$\ 1.36 \times 10^{14}$&$\ 1.205\times 10^{14}$&$\ 1.722\times 10^{14}$ \\
 \hline
\end{tabular}
\caption{Upper or lower bands on the parameters of the model for $\mathcal{P}_{\mathcal{R}} = 2.24 \times 10^{-9}$, $n_{s} = 0.9656$, $N_{e} = 50$ and $g_{\star} = 100$ using condition $T>H$ for four different $(p, q)$.}\label{l2}
\end{center}\end{table}

In figure \ref{ff2}, we have plotted the evolution of homogeneous dynamical quantities of the models,  namely $\frac{T}{M_{pl}}, \frac{\phi}{M_{pl}}, \frac{T}{H}, \frac{\rho_{R}}{V}$ and $Q_{G}$ versus number of e-folding. As we can observe, the Galileon dissipation factor starts from a small value roughly around $10^{-3}$ and becomes larger at the end of inflation in all panels.  In fact, the model  tends to start from a weak dissipation and ends up in a strong dissipation regime, as is expected in all the dissipating dynamical systems. Although, the rate of growth in the dissipation factor decreases by reducing the propagating speed of sound  (or increasing $q$) for fixed $p$, the dissipation ratio assumes larger values at the end of inflation by fixing $c_{s}$ and increasing $p$. Second, the radiation energy density is subleading at the beginning of inflation and becomes dominant (roughly around the half of inflaton energy density $\frac{\rho_{R}}{\rho_{\phi}} = \frac{1}{2}$) at the end of inflation in accordance with the Eq. (\ref{rr1}). Therefore, the system smoothly enters the radiation dominated epoch after inflation. Third, the temperature of the heat bath increases and the inflaton field value decreases as the number of e-folding decreases (or dissipation factor increases); therefore, the ratio of $\frac{T}{H}$ increases during inflation. In fact, the condition for being warm may easily be satisfied for larger values of the Galileon dissipation ratio as it is obvious from Eq. (\ref{tf}).

In figure \ref{ff3}, we have plotted log-log plot of the evolution of parameters versus $C_{T}$ where the minimum value of $C_{T}$ is derived from the condition $T>H$ and its maximum value is obtained for $\lambda \simeq 0.13$. There are several points regarding these plots which are worth mentioning; first, all panels unanimously show that the value of self-coupling of the Higgs self-interaction potential $\lambda$ increases and the coupling of Galileon self-interaction $M$ decreases by increasing the value of $C_{T}$ meaning that to have enough dissipation to keep thermal bath we need more light mediator fields coupled to inflaton during inflation. Therefore, the WGI scenario is within the limit of Planck data even for large $\lambda$, but we should accommodate a large number of fields to manage the dissipation process. Furthermore, the results we have obtained approach to that of the standard warm inflation \cite{Bastero-Gil:2016qru} ($C_{T}$ of order $10^{-2}$) for $\lambda$ of order $10^{-15}$. Second, the tensor-to-scalar ratio becomes larger by decreasing the value of $C_{T}$ and there is an upper bound on that where the condition $T>H$ is first met. Third, the field excursion is very small ($\Delta\phi \ll M_{pl}$) for large values of $C_{T}$ and becomes larger by decreasing the value of $C_{T}$. In fact, WGI scenario is a small field model for large values of $C_{T}$ and becomes a middle field model for small values of $C_{T}$ which is consistent with the results obtained in \cite{Visinelli:2016}.

In figure \ref{ff4}, we have shown the behavior of the tensor-to-scalar ratio as a function of the spectral index for different values of the e-folding number (or different values of dissipation factor) for the values of parameters shown in both Table \ref{l1} (Left panel) and Table \ref{l2} (Right panel). As is clear, the theoretical prediction of WGI shows very small values for tensor-to-scalar ratio of order $10^{-33}$ (Left panel) and of order $10^{-6}$ (Right panel) and therefore  is in excellent agreement with Plank data for $N_{e}=50$.  Finally, we have presented the values of parameters for four different models for $\lambda \simeq 0.13$ in Table \ref{l1} and for values where the condition $T>H$ is first met, in Table \ref{l2}. It should be noted that $Q_{G}, \frac{\rho_{R}}{V}$ and $\frac{\delta_{G\phi}}{\delta_{GX}}$ have same values in both tables since these parameters depend on $\zeta_{4}$ which do not change for fixed $\mathcal{P}_{\mathcal{R}}$ and $Q_{G}$. An interesting point in Table \ref{l1} and Table \ref{l2} is that the ratio $\frac{r}{n_{t}}$ is very small at the Hubble crossing time which confirms the results obtained for WI in \cite{Benetti:2017}, in other words, the amplitude of tensor perturbations in comparison to scalar perturbations is negligible in the WGI scenario.

\section{Summary and conclusions}\label{Conclusion}

Recently, a Warm G-inflation scenario \cite{Motaharfar:2017dxh} has been proposed in which the Galileon scalar field concurrently dissipates its energy to radiation field during inflation and therefore radiation becomes the dominant ingredient at the end of inflation and the universe smoothly crosses to a radiation dominated epoch. The most absorbing feature of the WGI model is that it may obtain large self-coupling, $\lambda \sim 0.13$, in agreement with quantum theoretical predictions solving the model for a Higgs self interaction potential due to the presence of new degrees of freedom coming from dissipation process and Galileon self-interaction, in contrast to its counterpart in cold scenarios which suffer from lack of oscillatory phase during reheating. In \cite{Motaharfar:2017dxh}, the authors calculated the power spectrum of the WGI scenario  in a weak dissipation regime or for a dissipative coefficient which is independent of temperature ($c=0$) while, as the authors in \cite{Graham:2009} have shown, the power spectrum may get modified with a growing mode function in high dissipation regime for a temperature dependence dissipative coefficient ($c \neq 0$) due to the coupling between inflaton and radiation perturbed field equations. As a result, such growing mode may modify curvature power spectrum in such a way as to significantly change the constraints on the parameters of the model.

Keeping that in mind, we have investigated how the  backreaction of the produced radiation leading to a growing mode in the inflaton's perturbations can be controlled by decreasing the propagating sound speed of the inflaton's perturbations in a WGI scenario. In fact, the propagating sound speed $c_{s}$ is translated to perturbed radiation equation in such a way that produces a damping effect in radiation perturbations. This is expected when the propagating sound speed of perturbations of the Galileon inflaton is less than the standard WI (in fact, the Hubble radius is smaller than that of the Standard WI scenario) where the resulting radiation has less time to equilibrate in the radiation bath, leaving significant imprint on perturbations of the inflaton on Hubble crossing time since its approach to equilibrium is controlled by the propagating speed of sound $c_{s}$. Therefore, we showed that decreasing the propagating sound speed of perturbations effectively damps the radiation fluctuations so as to avoid altogether the appearance of a growth mode in the resulting perturbations. The results we have obtained are model independent and that the overall effect of the compensation of the growing mode depends on $c_{s}$. Therefore the resulting power spectrum for $(c\neq 0)$ asymptotically approaches the power spectrum obtained for $(c=0)$ and the growth mode completely disappears for $c_{s}\ll 1$.

The rest of the paper was devoted to constrain the parameters of the model with Higgs self interaction potential and linear temperature dependence dissipative coefficient using the Planck 2015 likelihood. To this end, we showed that the number of e-folding can be obtained as a function of the newly defined Galileon dissipation factor $Q_{G}$ which is usable for both weak and strong dissipation regimes and all parameters of the model can also be written in  terms of $Q_{G}$. Consequently,  the model tends to start with a small Galileon dissipation factor of order $10^{-3}$ and move to high Galileon dissipation regime at the end of inflation and therefore, the power spectrum is completely consistent with the observational data at large scales and becoms amplified (or becomes more blue-tilted) at small scales. Although, as was mentioned earlier, such a growing mode of the power spectrum may be controlled by decreasing the value of the speed of sound  $c_{s}$. Based on our analyses, we found that having large self-coupling to be consistent with quantum field theoretical predictions for Higgs self-interaction potential requires to accommodate a large value for $C_{T}$ which means that many light mediator fields coupled to the inflaton field are required in order to produce enough dissipation as we expected. Although, we showed that there is a monotonic relation between $C_{T}$ and $\lambda$ whereby our results coincides with the results obtained in \cite{Bastero-Gil:2016qru} ($C_{T}$  $\sim 10^{-2}$) for smaller self-coupling of the Higgs potential ($\lambda \sim 10^{-15}$).

Finally, we emphasize that such damping effects of the propagating speed of sound will also have an impact on the evolution of second-order perturbations and the resultant non-Gaussianity. The upcoming cosmological data are expected very soon to put tight constraint on the level of non-Gaussianity of primordial spectrum which will obviously help to distinguish different inflationary models. Warm inflation is categorized as a model with non-negligible non-linearity parameter $f_{NL}$ for non-Gaussianity due to its multi-field nature. This parameter has been computed for both $T$ independent \cite{Moss:2007cv} and $T$ dependent \cite{Moss:2011qc} dissipative coefficient in WI scenarios where the latter provides an extra non-linear source in the second-order equation,  resulting in larger non-Gaussianity for larger dissipation ratio. However, if the coupling between first order perturbations of the inflaton and radiation fields is suppressed by a decreasing propagating sound speed,  we expect qualitatively that the same occurs at the second order in the WGI scenario. Therefore, we hope to study this effect and other possible issues in a separate work in the near future.

\appendix
\section{Details}\label{we}

\subsection{Compelete SLE for WGI}

Second-order Langevin equation for WGI in complete form reads as

\begin{align}
 \bold{B} \ddot\Phi ({\bf x},t)+ 3 H \bold{A} \dot\Phi ({\bf x},t) + V_{,\Phi} (\Phi) -\bold{F} \frac{\nabla^{2}}{a^{2}}\Phi(x, t) + \bold{K}\notag = \xi(x,t),
\end{align}
where

\begin{align}
\nonumber\bold{A}&= 1+Q + 3H\dot\Phi G_{,X}+ \frac{\dot{H}}{H}\dot\Phi G_{,X}- 2G_{,\Phi}+2X G_{,\Phi X}
- \frac{\dot\Phi}{3H}G_{,\Phi\Phi}
-\frac4{3H}G_{,\Phi X}\frac{{\nabla}\dot\Phi.{\nabla}\Phi}{a^2} + \frac23 G_{,\Phi X} \frac{(\nabla \Phi)^{2}}{a^{2}}\\&\notag + H \dot\Phi G_{,XX}\frac{(\nabla\Phi)^{2}}{a^2} -2\dot\Phi G_{,XX}\frac{{\nabla}\dot\Phi.{\nabla}\Phi}{a^2} + \frac{\dot\Phi}{3H} G_{,XX}\frac{(\nabla \dot\Phi)^{2}}{a^{2}}+ G_{,XX} \frac{(\nabla \Phi.\nabla)^{2}\Phi}{a^4} \notag + \frac{2}{3H} G_{,XX} \frac{\nabla^{2}\Phi (\nabla\dot\Phi.\nabla\Phi)}{a^{4}} \\&-\frac{1}{3H} G_{,XX} \frac{\nabla^{2}\Phi (\nabla\Phi)^{2}}{a^{4}}  - \frac{1}{3H} G_{,XX} \frac{(\nabla \dot\Phi.\nabla)(\nabla \Phi.\nabla\Phi)}{a^4}\\ \nonumber
\bold{B}&= 1-2G_{,\Phi}-2X G_{,\Phi X}+6H\dot\Phi G_{,X}+ 6H\dot\Phi X G_{,XX}
-2\left(G_{,X}+X G_{,XX}\right)\frac{\nabla^2\Phi}{a^2}
-H\dot\Phi G_{,XX}\frac{(\nabla\Phi)^{2}}{a^2} \notag\\&+ \frac{1}{2}G_{,XX}\frac{(\nabla \Phi.\nabla)^{2}\Phi}{a^4} \\ \nonumber
\\
\bold{F} & = 1- 2 G_{,\Phi} + 2XG_{,\Phi X} + 4 H \dot\Phi G_{,X}, \\
\notag\bold{K}& =  -(H^2+\dot{H})G_{,X}\frac{(\nabla\Phi)^{2}}{a^2} - 4 H G_{,X} \frac{\nabla\dot\Phi.\nabla\Phi}{a^{2}}-2 G_{,X} \frac{(\nabla\dot\Phi)^{2}}{a^{2}} \notag -H^{2}G_{,XX}\frac{(\nabla\Phi)^{4}}{a^{4}}- G_{,XX} \frac{(\nabla\dot\Phi.\nabla\Phi)^{2}}{a^{4}} \\& \notag-\frac16 G_{,X} \frac{(\nabla \times \nabla\Phi)^{2}}{a^{4}}+ G_{,X} \left(\frac{\nabla^{2}\Phi}{a^{2}}\right)^{2} - G_{,\Phi X} \frac{(\nabla\Phi.\nabla)^{2}\Phi}{a^{4}} + 2HG_{,XX} \frac{(\nabla\Phi.\nabla\Phi)(\nabla\dot\Phi.\nabla\Phi)}{a^{4}}\\& -\frac{1}{2}G_{,XX}\frac{\nabla^{2}\Phi (\nabla\Phi.\nabla)^{2}\Phi}{a^6}+ \frac{1}{12}G_{,XX} \frac{\left((\nabla\times\nabla\Phi)\times\nabla\Phi\right)^{2}}{a^{6}}.
\end{align}

\subsection{Perturbed SLE for WGI}

To perturb Eq. (\ref{SLE}) around its homogeneous inflaton field up to first order in perturbations we need

\begin{align}
&X \rightarrow X +\delta X,\\
&\bold{A}\rightarrow\mathcal{A}+\delta \bold{A},\\
&\bold{B}\rightarrow \mathcal{B}+\delta \bold{B},\\
&\bold{F} \rightarrow \mathcal{F} + \delta \bold{F},\\
&\bold{K} \rightarrow \mathcal{K} + \delta \bold{K},\\
&G(\Phi, X)\simeq G(\phi, X)+G_{,X}\delta X, \label{fv}\\
&V(\Phi)\simeq V(\phi)+V_{,\phi}\delta\phi,
\end{align}
where we have dropped $G_{,\phi} \delta \phi$ in Eq. (\ref{fv}) since it is  second order in perturbations and to derive the spectral index we just need first order perturbed SLE and

\begin{align}
& \delta X = \dot\phi \delta\dot\phi,\\
&\delta \bold{A}\simeq \delta Q+3H(G_{,X}+2XG_{,XX})\delta\dot\phi,\\
&\delta \bold{B}\simeq 6H(G_{,X}+5XG_{,XX}+2X^2G_{,XXX})\delta\dot\phi-2(G_{,X}+XG_{,XX})\frac{\nabla^2}{a^2}\delta\phi,\\
& \delta \bold{F} \simeq 4H(G_{,X}+ 2XG_{,XX})\delta\dot\phi,\\
& \mathcal{F} \simeq 1+4 H \dot \phi G_{,X}.
\end{align}
Therefore, the SLE can be expanded as follows

\begin{align}
\left(\mathcal{B}+\delta \bold{B}\right)\left(\ddot\phi+\delta \ddot\phi\right) + 3 H \left(\mathcal{A}+\delta \bold{A}\right)\left( \dot\phi+\delta \dot\phi\right) - \left(\mathcal{F}+ \delta \bold{F}\right) \frac{\nabla^{2}}{a^{2}}\delta\phi + V_{,\phi}+V_{,\phi\phi}\delta\phi \simeq \xi(x,t).
\end{align}
Utilizing Eq. (\ref{d1}) and dropping second-order perturbation terms results in the following perturbed SLE

\begin{align}
\mathcal{B}\delta \ddot\phi + 3 H \mathcal{A} \delta \dot\phi+3H\delta \bold{A}\dot\phi - \mathcal{F} \frac{\nabla^{2}}{a^{2}}\delta\phi+V_{,\phi\phi}\delta\phi \simeq \xi(x,t),
\end{align}
where we have dropped $\ddot\phi \delta B$ since it is second order in perturbations

\begin{align}
\ddot\phi \delta \bold{B} = \frac{\ddot\phi}{H\dot\phi} H\dot\phi \delta \bold{B} = \delta_{\phi} \left[6H\frac{\delta_{GX}}{\delta_{X}}\left(1+ 5 \kappa_{X}+ 2 \kappa_{XX}\right)\delta\dot\phi- 2\frac{\delta_{GX}}{\delta_{X}}(1+ \kappa_{X})\frac{\nabla^{2}}{a^{2}}\delta\phi \right] \sim \epsilon^{2},
\end{align}
with $\kappa_{XX} = \frac{X^{2}G_{,XXX}}{G_{,X}}$. Hence, the SLE can be written in Fourier space as

\begin{align}
\mathcal{B}\delta \ddot\phi + 3 H \left(Q+\mathcal{B}\right) \delta \dot\phi +\left( k^{2}a^{-2}\mathcal{F} +V_{,\phi\phi}\right)\delta\phi + 3 H \dot \phi \delta Q \simeq  \xi(\bold{k},t)\label{er},
\end{align}
where

\begin{align}\label{ap1}
\notag\\ 3 H \dot \phi \delta Q = \delta\Gamma\dot\phi=&\Gamma c \frac{\delta T}{T}\dot\phi + \Gamma_{,\phi} \dot\phi \delta\phi
\notag\\ = & 3cH(\frac{\Gamma}{3H})\frac{\delta\rho_r}{4\rho_r}\dot\phi - 3 H^{2} Q \frac{\beta}{\mathcal{A}} \delta\phi
\notag\\ = & 3cH(\frac{\Gamma}{3H})\frac{\delta\rho_r}{\Gamma\dot\phi^2/H}\dot\phi- 3 H^{2}Q \frac{\beta}{\mathcal{A}} \delta\phi
\notag\\ = & 3cH^2Q(\Gamma\dot\phi)^{-1}\delta\rho_r- 3 H^{2}Q \frac{\beta}{\mathcal{A}} \delta\phi.
\end{align}
Thus, inserting (\ref{ap1}) into (\ref{er}) and using the propagating sound speed $c_{s} = \frac{\mathcal{F}}{\mathcal{B}}$, we obtain Eq. (\ref{m1}).


\end{document}